\documentclass[conference,compsoc]{IEEEtran}
\IEEEoverridecommandlockouts
\usepackage{hyperref}       
\usepackage{url}            
\usepackage{amsfonts}       
\usepackage{graphicx}       
\usepackage{amsthm}
\usepackage{amsmath}
\usepackage{algpseudocode, algorithm}
\usepackage{subcaption}
\usepackage{diagbox}
\usepackage[table]{xcolor}
\usepackage[nowidering]{yhmath}
\usepackage{threeparttable}
\usepackage{adjustbox}
\usepackage{mathtools}
\usepackage{makecell}
\usepackage{slashbox}
\usepackage{xr}
\makeatletter
\newcommand*{\addFileDependency}[1]{
  \typeout{(#1)}
  \@addtofilelist{#1}
  \IfFileExists{#1}{}{\typeout{No file #1.}}
}
\makeatother

\algnewcommand\INPUT{\item[\textbf{Input:}]}%
\algnewcommand\OUTPUT{\item[\textbf{Output:}]}%

\algdef{SE}
[STRUCT]
{Struct}
{EndStruct}
[1]
{\textbf{struct} \textsc{#1}}
{\textbf{end struct}}

\ifCLASSOPTIONcompsoc
  \usepackage[nocompress]{cite}
\else
  \usepackage{cite}
\fi
\ifCLASSINFOpdf
\else
\fi
\begin{document}
\newtheorem{theorem}{Theorem}[section]
\newtheorem{lemma}{Lemma}[section]
\newtheorem{definition}{Definiton}[section]
\newtheorem{corollary}{Corollary}[section]
\newtheorem{claim}{Claim}[section]
\newtheorem{proposition}{Proposition}[section]
\newtheorem{conjecture}{Conjecture}[section]
\newtheorem{property}{Property}[section]
\newtheorem{example}{Example}
\newtheorem{assumption}{Assumption}

%
\title{Improving Count-Mean Sketch as the Leading Locally Differentially Private Frequency Estimator for Large Dictionaries}

\author{\IEEEauthorblockN{Mingen Pan}
\IEEEauthorblockA{
Independent Researcher \IEEEauthorrefmark{1} \thanks{\IEEEauthorrefmark{1} Currently working at Google LLC. This work was conducted independently and does not reflect the views or endorsement of Google.} \\
Email: mepan94@gmail.com}
}


%


\maketitle

\begin{abstract}

This paper identifies that a group of latest locally-differentially-private (LDP) algorithms for frequency estimation, including all the Hadamard-matrix-based algorithms, are equivalent to the private Count-Mean Sketch (CMS) algorithm with different parameters. Therefore, we revisit the private CMS, correct errors in the original CMS paper regarding expectation and variance, modify the CMS implementation to eliminate existing bias, and optimize CMS using randomized response (RR) as the perturbation method. The optimized CMS with RR is shown to outperform CMS variants with other known perturbations in reducing the worst-case mean squared error (MSE), $l_1$ loss, and $l_2$ loss. Additionally, we prove that pairwise-independent hashing is sufficient for CMS, reducing its communication cost to the logarithm of the cardinality of all possible values (i.e., a dictionary). As a result, the optimized CMS with RR is proven theoretically and empirically as the leading algorithm for reducing the aforementioned loss functions when dealing with a very large dictionary. Furthermore, we demonstrate that randomness is necessary to ensure the correctness of CMS, and the communication cost of CMS, though low, is unavoidable despite the randomness being public or private.

\end{abstract}


%
\IEEEpeerreviewmaketitle

\section{Introduction}

Frequency is a fundamental metric to describe the properties of a dataset, counting the occurrences of specific values. Calculating the frequency of each value is straightforward if the original values are known to the analyst, as one can generate a hash map to record the occurrences of all values in the dataset. This method has linear time and space complexity relative to the dataset size. However, frequency calculation becomes challenging when data is protected, and their values are not directly accessible to the analyst.

One of the most popular approaches to protect data is differential privacy (DP) \cite{dwork2006calibrating}, which is considered the \textit{de facto} standard for extracting statistics from a dataset while preserving individual data privacy. DP introduces sufficient noise to the statistical outputs such that the participation of any individual's data cannot be confidently inferred from the query results. However, DP still requires individuals to upload their data to a server before extracting statistics, posing risks of eavesdropping during data transmission or data breaches on the server side \cite{hassan2019breach}. To address these concerns, Local Differential Privacy (LDP) \cite{LDP} has been proposed, which requires enough perturbation on individual data on the client side before sending them to the server for statistical analysis, theoretically preventing eavesdropping or data breaches. However, LDP also adds more challenges to calculating statistics from a dataset, including frequency estimation.


A major challenge of LDP-preserving frequency estimation is that traditional mechanisms, such as randomized response \cite{wang2016using}, result in the precision of the estimated frequency increasing linearly with the number of all possible values in a dataset (referred to as a dictionary), making the estimation infeasible if the dictionary is too large. Several algorithms \cite{bassily2015hadamard, kairouz2016rappor, apple_privacy, bassily2017countsketch, wang2017localhashing, ye2018subset, acharya2019hr, chen2020rhr, pagh2022countsketch} have been developed to address this problem, and their precision is proven to be independent of the dictionary size, though some of them still have communication cost linear to the dictionary size. These algorithms will be further introduced in the Previous Work (Section \ref{sec:previous_work}). Many of these algorithms (e.g., \cite{bassily2015hadamard, wang2017localhashing, acharya2019public_hr, chen2020rhr}) are found to be equivalent to the Count-Mean Sketch (CMS) algorithm (see Section \ref{sec:connection_with_existing} for more details). CMS hashes the original values through randomly selected hash functions and estimates the original frequencies based on the frequencies of the hashed values. The LDP version of CMS was originally developed by \cite{apple_privacy}, but it had minor issues in calculating the expectation and variance (see Section \ref{sec:correct_original_cms}) and did not explore the optimized parameters for CMS. Therefore, our work provides a corrected calculation of expectation and variance (Sections \ref{sec:expectation} and \ref{sec:variance}) and optimize CMS using randomized response as the perturbation method (referred to as OCMS+RR). OCMS+RR is shown to outperform CMS variants with other known perturbations in reducing the worst-case mean squared error ($\widehat{MSE}$), $l_1$ loss, and $l_2$ loss (Section \ref{sec:optimized_cms}). Moreover, our work proves that pairwise-independent hash functions are sufficient for CMS, and Section \ref{sec:imperfect_hashing} proposes a practical approach to construct a pairwise-independent hashing family given any range of a hash function. The pairwise independence can reduce the communication cost to the logarithm of the dictionary size, allowing OCMS+RR to be applied to datasets with very large dictionaries. The pseudocode of OCMS+RR is presented in Section \ref{sec:implementation}  (source code available at \url{https://github.com/mingen-pan/CountMeanSketch}). When the dictionary size is much larger than the privacy factor $\epsilon$ defined by LDP, the aforementioned loss functions (defined in Section \ref{sec:frequency_estimation}) are reduced by OCMS+RR as 
\begin{align*}
      & \widehat{MSE} \le \frac{e^{\epsilon/2}}{n(e^{\epsilon/2} - 1)^2} 
\end{align*}
\begin{align*}
    & l_1  \xrightarrow{d \gg e^{\epsilon}} \frac{2d e^{\epsilon / 2}}{\sqrt{n} (e^{\epsilon} - 1)} \\
    & l_2  \xrightarrow{d \gg e^{\epsilon}} \frac{4 d e^{\epsilon}}{n (e^{\epsilon} - 1)^2} 
\end{align*}

\noindent where $n$, and $d$ are the dataset size and dictionary size, respectively. As a result, Section \ref{sec:compare} compares existing algorithms with OCMS+RR and concludes that our OCMS+RR is the leading algorithm for reducing the worst-case MSE ($\widehat{MSE}$), $l_1$ loss, and $l_2$ loss, while its communication cost is only logarithmic relative to the dictionary size. Additionally, Section \ref{sec:necessity_of_randomness} proves the necessity of randomness in CMS and provides a counterexample to an attempt to de-randomize a CMequivalent algorithm by \cite{acharya2019public_hr}. Section \ref{sec:public_private_randomness} shows that the communication cost of CMS is unavoidable, despite the randomness being public or private. Section \ref{sec:experiment} empirically demonstrates that OCMS+RR outperforms other algorithms. 

\subsection{Previous Work}
\label{sec:previous_work}

RAPPOR \cite{kairouz2016rappor} is the first algorithm to estimate frequency with precision independent of dictionary size. It represents the original value as a one-hot vector and applies LDP perturbation to every bit of the array. However, the communication cost is proportional to the dictionary size, as the one-hot vector is the same size as the dictionary. An alternative version of RAPPOR \cite{rappor} hashes an original value into a smaller range than the dictionary size, resulting in a much smaller one-hot encoded vector. This version uses gradient descent to search for one possible solution for the original frequencies. However, this mapping is equivalent to dimension reduction in linear algebra, so the found solution is not guaranteed to be close to the original frequencies. To tackle the large dictionary, an algorithm was developed by \cite{bassily2015hadamard}, which transforms an original value to a column of a Hadamard matrix, uniformly samples one element from the column, and perturbs it. The original frequencies can be reconstructed from the transformed values by applying the Hadamard matrix again. The precision of this approach is independent of the dictionary size, and the communication cost is the logarithm of the dictionary size. The Apple Privacy team \cite{apple_privacy} modified the Count-mean Sketch (CMS) algorithm to be LDP. However, they failed to derive the expectation and variance correctly (see Section \ref{sec:correct_original_cms}) and did not explore the optimized settings of CMS. In addition to CMS, other members of the Count Sketch family have also been adapted to LDP (e.g., \cite{bassily2017countsketch, pagh2022countsketch}). Wang et al.  \cite{wang2017localhashing} summarized existing approaches and proposed a variant of RAPPOR and a local hashing method, which were claimed to be optimal in terms of MSE. However, \cite{wang2017localhashing} assumed the estimated frequencies to be always close to zero, so their optimality did not hold when the original frequency approaches one. Simultaneously,  the Subset Selection algorithm \cite{ye2018subset} was proposed to estimate frequency with leading precision in $l_1$ / $l_2$ losses, but its communication cost is also proportional to the dictionary size when the privacy requirement is high. Acharya et al. \cite{acharya2019hr} later integrated the Hadamard matrix into the subset selection and reduced the communication cost to the logarithm of the dictionary size. This method was further improved by \cite{chen2020rhr}, which applies the Hadamard transform twice. Though its performance can achieve the state-of-the-art order of precision in $l_1$ / $l_2$ losses, it may perform worse than the Hadamard Response in terms of MSE (as discussed in Section \ref{sec:rhr}).

\section{Background}
\subsection{Local Differential Privacy (LDP)}
\label{sec:ldp}
Local Differential Privacy (LDP) was initially proposed in \cite{LDP} to ensure that querying an object does not significantly disclose its actual value. Consider a variable $V$ that is perturbed by a mechanism $M$:

\begin{definition}
   $M$ is $\epsilon$-LDP if and only if $\forall v, v' \in \mathcal{V}$, and $\forall u \in \mathcal{U}$,
\begin{equation}
    Pr(M(V) = u | V = v) \le e^{\epsilon} Pr(M(V) = u | V = v') , \notag
\end{equation} 
\end{definition}

\noindent where $\mathcal{V}$ and $\mathcal{U}$ represent the dictionary (all possible values) of $V$ and the range of $M$, respectively. This paper may abbreviate $\left( M \left( V \right) | V = v \right)$ as $M(v)$ when the variable is not the focus and its value is known. 

Given the above definition, $P$ is defined as a $|\mathcal{U}| \times |\mathcal{V}|$ matrix, where $P[u, v] = Pr(M(v) = u)$. Here $u$ and $v$ are represented by an index between $1..|\mathcal{U}|$ and $1..|\mathcal{V}|$, respectively. We consider only cases where $|\mathcal{U}| \ge |\mathcal{V}|$ and $P$ is full rank, i.e., $\text{rank}(P) = |\mathcal{V}|$. This condition should hold for all meaningful LDP mechanisms, as the expected frequencies of the perturbed values correspond to a linear transformation of the original frequencies, and a rank-deficient transformation would prevent the reconstruction of the original frequencies from the perturbed values. Widely-used algorithms (e.g., \cite{wang2016using, kairouz2016rappor}) satisfy this condition. Subsequently, a matrix $Q$ is constructed as $(P^T P)^{-1} P^T$, which satisfies $I = QP$ with $I$ being an identity matrix. Given an output $u$ from $M$, its decoding is defined to output $Q[:, u]$, i.e., the $u$-th column of $Q$. Thus, the reconstruction process (perturbation and decoding) of a variable $V$ is defined as $R(V) = Q[:, M(V)]$. Similarly, $\left( R \left( V \right) | V = v \right)$ is abbreviated as $R(v)$ when the variable is not the focus and its value is known. An example of the reconstruction is presented in Appendix \ref{sec:example_reconstruction}. The randomness of $R(v)$ comes from $M(v)$, and each possible value of $R(v)$ is a vector of size $|\mathcal{V}|$ satisfying:

\begin{property}
\label{pro:ldp_1}
    \begin{equation*}
        \begin{gathered}
             E[R(v)[v]] = 1  \\
             \forall v' \ne v: E[R(v)[v']] = 0
        \end{gathered} 
    \end{equation*}
    \noindent where $E$ denotes expectation.
\end{property}

\noindent Proof: represent any $u$ from $M(v)$ as a one-hot vector of size $|\mathcal{U}|$, where the $u$-th element is one. A random variable $\mathbf{u}$ represents all the possible outputs from $M(v)$ as one-hot vectors. Subsequently, $R(v) = Q \mathbf{u}$. Therefore, $E[R(v)] = Q E[\mathbf{u}] = Q P \mathbf{e}_v = \mathbf{e}_v$, where $\mathbf{e}_v$ is a one-hot vector with the $v$-th element being one. $\qed$

Meanwhile, we require the LDP mechanism $M$ to be stateless, meaning that its output depends only on its input. This is formulated as
\begin{equation}
    Pr(M(V) = u | V = v) = Pr(M(V) = u | V = v, \theta) , \notag
\end{equation}

\noindent where $\theta$ represents any other information (except the randomness generator used by $M$). Considering two variables $V$ and $V'$, whose values are determined prior to perturbation, the stateless requirement ensures that their reconstruction process $R(V)$ and $R(V')$ are mutually independent random variables, formulated as

\begin{property}
The covariance of $R(V)$ and $R(V')$ satisfies $\forall v, v', v_a, v_b \in \mathcal{V} : Cov( R( V ) [v_a] , R( V' ) [v_b] | V = v, V' = v') = 0$.
\label{pro:ldp_2}
\end{property}

Widely-used algorithms (e.g., \cite{wang2016using, kairouz2016rappor}) satisfy this property. Additionally, we observe that these LDP mechanisms also exhibit symmetry in variance, denoted as $Var$, formulated as:

\begin{property}
 Every element of $\{Var(R(v)[v']) | v, v' \in \mathcal{V}, v \ne v' \}$ has an identical value, denoted as $Var(R | \ne)$; every element of $\{Var(R(v)[v]) | v \in \mathcal{V} \}$ has an identical value, denoted as $Var(R | =)$.
\label{pro:ldp_3}
\end{property}

\subsection{Frequency Estimation}
\label{sec:frequency_estimation}
Let's define some symbols: $X^{(i)}$ is the $i$-th object of an arbitrary dataset of size $n$; the possible values of all the objects comprise a dictionary $[d]$, where $d$ is its size, and $\forall i \in [n]: X^{(i)} \in [d]$. The frequency of value $x$ is defined as $f(x) = \frac{\sum_{i \in [n]} \mathbf{1}\{ X^{(i)} = x\}}{n}$. The LDP workflow of frequency estimation typically proceeds as follows: \vspace{2pt} 

1. Each object $X^{(i)}$ is perturbed by an LDP mechanism, which outputs $\tilde{X}^{(i)}$.

2. An analyst collects and aggregates the outputs as $\hat{f}(x) = A_x([\tilde{X}^{(i)}]_{i \in [n]})$, if the analyst is interested in $f(x)$, where $A_x$ denotes the aggregation. The aggregation can be performed for every $x \in [d]$.

\vspace{2pt} A potential scenario involves a group of clients sending their perturbed responses to a server, while their unperturbed responses constitute a dataset. The server then estimates the frequency of the unperturbed responses.

$\hat{f}(x)$ in Step 2 is an estimator of $f(x)$. Existing literature usually uses $l_1$ and $l_2$ losses to measure its overall precision, which are defined as follows:

\begin{equation*}
    \begin{gathered}
        l_1(\hat{f}) = \max_{\forall \mathbf{X}: |\mathbf{X}| = n} E[ \sum_{x \in [d]} |\hat{f}(x) - f(x)| ]\\
        l_2(\hat{f}) = \max_{\forall \mathbf{X}: |\mathbf{X}| = n} E[ \sum_{x \in [d]} (\hat{f}(x) - f(x))^2 ]
    \end{gathered} 
\end{equation*}

\noindent where $\mathbf{X} = [X^{(i)}]_{i \in [n]}$, and $\forall \mathbf{X}$ assumes the size $n$ is constant. These two losses measure the overall precision of the estimation. However, in many cases, an analyst is also interested in the precision of the frequency estimation of an individual value. The mean squared error (MSE), formulated as $MSE(\hat{f}(x))=E[(\hat{f}(x) - f(x))^2]$, is commonly used to measure the precision of an estimation, reflecting both the variance and bias of the estimation. Since an analyst is assumed to have no knowledge of the actual dataset, they need to assume the worst case when calculating the MSE, which is defined as

\begin{equation}
    \widehat{MSE}(\hat{f})=  \max_{\forall \mathbf{X}: |\mathbf{X}| = n} \max_{x \in \mathbf{x}} MSE(\hat{f}(x)),
\end{equation}

\noindent where $\widehat{MSE}$ denotes the upper bound of MSE, i.e., the worst case; $\mathbf{x}$ denotes the values of interest, which is $[d]$ by default. If the estimator $\hat{f}$ is unbiased, the worst-case MSE is the same as the worst-case variance, which is also positively correlated to the confidence interval (CI). Thus, an unbiased estimator optimized for $\widehat{MSE}$ is also optimized for CI.

\section{Revisit Count-mean Sketch}
\label{sec:revisit_cms}

\begin{figure}[htp!]
\centerline{\includegraphics[width=1.0\columnwidth]{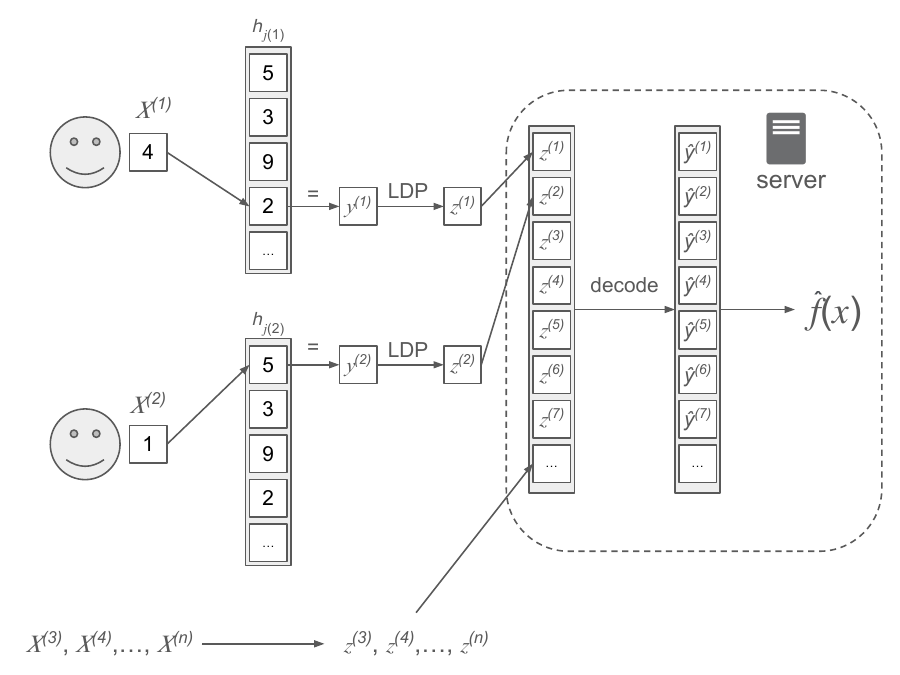}}
\caption{Visualization of Count-mean Sketch.}
\label{fig:cms}
\end{figure}
Count-Mean Sketch (CMS) is a probabilistic data structure for frequency estimation, which (1) uses multiple hash functions to map objects into different counters, and (2) calibrates the bias in each counter to estimate the frequency of the values of interest. Suppose there is a dataset $[X^{(i)}]_{i \in [n]}$, each with a value belonging to a dictionary $[d]$. When setting up the CMS program, a hashing family $\mathcal{H}$ with  $k$ hash functions mapping from $[d]$ to $[m]$ is provided. When uniformly sampling a hash function $h$ from $\mathcal{H}$, we expect that $\forall (x_1, x_2) \in [d]^2: Pr(h(x_1) = h(x_2)) \approx \frac{1}{m}$, which will be further discussed in Section \ref{sec:expectation}. CMS operates with the following steps (Fig. \ref{fig:cms}):

1. Each object $X^{(i)}$ is assigned a hash function uniformly sampled from $\mathcal{H}$. The assigned hash function is denoted as $h_{j^{(i)}}$, where $j^{(i)} \in [k]$ is the index of the hash function. The server and the client sync on the hash-function assignment.

2. Hash $X^{(i)}$ as $y^{(i)} = h_{j^{(i)}}(X^{(i)}) \in [m]$. Then, $y^{(i)}$ is perturbed by an $\epsilon$-LDP mechanism as $z^{(i)}$. 

3. Collect all the responses $(z^{(i)})_{i \in [n]}$, decode them following Section \ref{sec:ldp} as $(\hat{y}^{(i)})_{i \in [n]}$, where each $\hat{y}^{(i)}$ is an estimator of $y^{(i)}$, i.e., $\hat{y}^{(i)} = R(y^{(i)})$, with $R$ defined in Section \ref{sec:ldp}. Notice that each $\hat{y}^{(i)}$ is a vector of size $m$. 

4. The frequency of any value $x \in [d]$ can be estimated as
\begin{equation}
    \hat{f}(x) = \frac{m}{ n (m - 1) } \bigl[ \sum_{i \in [n]}  \hat{y}^{(i)}[h_{j^{(i)}}(x)] \bigr]  - \frac{1}{m - 1}. \label{eq:estimator_def}
\end{equation}

Notably, $\hat{f}(x)$ may produce values outside the interval $[0, 1]$ due to perturbations in the collected data. One straightforward solution is to truncate $\hat{f}(x)$ to $[0, 1]$, i.e. $\min(1, \max(0, \hat{f}(x)))$, which also improves precision by eliminating deviations beyond the bounds. More advanced algorithms, such as projection onto a simplex, can also be used to enforce $\hat{f}(x) \in [0, 1]$ and improve precision, as discussed in \cite{kairouz2016rappor}.

Regarding privacy guarantee, the only output that depends on the value of $X^{(i)}$ is $z^{(i)}$, which has been perturbed by an $\epsilon$-LDP mechanism. The postprocessing of the outputs from an LDP mechanism does not change its privacy guarantee \cite{dwork2014algorithmic}, so we have proved:

\begin{theorem}
    The CMS frequency estimator $\hat{f}(x)$ is $\epsilon$-LDP.
\end{theorem}

\subsection{Expectation and its Bias}
\label{sec:expectation}

This section derives the expectation of $\hat{f}(x)$. Here, we only consider the randomness of assigning hash functions to objects and the LDP perturbation, and regard $\mathcal{H}$ and $[X^{(i)}]_{i \in [n]}$ as constant. That is, $h_j(x)$ will not be considered a random variable; $\forall j \in [k], \forall (x, x') \in [d]^2$, whether $h_j(x') = h_j(x)$ is determined in advance as well, which will be denoted as $c_j(x, x')$.

The hashing family $\mathcal{H}$ is treated as constant for the following reasons: (1) $\mathcal{H}$ is visible prior to executing CMS, as the server must sync it with all clients in advance to ensure that the clients know which hashing family to sample; and (2) though the hashing family $\mathcal{H}$ can be regenerated before each run of the CMS, we will show later that it is sufficient to have a constant hashing family $\mathcal{H}$ at all times.

Some literature \cite{acharya2019public_hr} also proposed a deterministic approach to assign hash functions, which will be discussed in Section \ref{sec:necessity_of_randomness}. Here, we only focus on the CMS defined above and require the hash-function assignments to be random events. Now, the expectation of the frequency estimation by CMS is derived as:

\begin{theorem}
    The expectation of $\hat{f}(x)$ is
    \begin{multline}
        E[\hat{f}(x)] = \frac{m}{m - 1} [f(x) \\ + \sum_{x' \in [d] \backslash x} \sum_{j \in [k]}  \frac{c_j(x, x')}{k} f(x')] - \frac{1}{m - 1} 
        \label{eq:expect_fx}
    \end{multline}

    \noindent where $f(x)$ is the frequency of value $x$; $[d] \backslash x$ represents the set of $[d]$ excluding $x$.
    \label{thm:expect_fx}
\end{theorem}

\begin{corollary}
\label{cor:fx_no_bias}
   If and only if $\sum_{j \in [k]} \frac{c_j(x, x')}{k} = \frac{1}{m}$ is satisfied for all $x \ne x'$, then
   \begin{equation*}
        E[\hat{f}(x)] = f(x) .
   \end{equation*} 
\end{corollary}

\noindent The proof of Theorem \ref{thm:expect_fx} and Corollary \ref{cor:fx_no_bias} is presented in Appendix \ref{sec:expect_proof}. It is notable that $\forall x \ne x': \sum_{j \in [k]} \frac{c_j(x, x')}{k} = \frac{1}{m}$ is equivalent to $\Pr_{h \in \mathcal{H}}(h(x) = h(x')) = \frac{1}{m}$, where $h \in \mathcal{H}$ represents uniform sampling, which is also the definition of a pairwise-independent hashing family. Thus, Corollary \ref{cor:fx_no_bias} can be rephrased as:

\begin{corollary}
    If and only if $\mathcal{H}$ is pairwise independent, i.e., $\Pr_{h \in \mathcal{H}}(h(x) = h(x')) = \frac{1}{m}$, we have $E[\hat{f}(x)] = f(x)$.
    \label{cor:pairwise_unbiased}
\end{corollary}

\subsection{Variance}
\label{sec:variance}

Assuming that the LDP reconstruction process $R$ in Step 3 of CMS satisfies Properties \ref{pro:ldp_1} and \ref{pro:ldp_2}, we can derive:

\begin{lemma}
If $R$ satisfies Properties \ref{pro:ldp_1} and \ref{pro:ldp_2}, we have
    \begin{multline}
        Var( \hat{f}(x) ) = \frac{m^2}{(m - 1)^2 n} \sum_{x' \in [d]} \biggl[ \biggl( \frac{1}{k} \sum_{j \in [k]}  Var\bigl( \\  R(h_j(x')) [h_{j}(x)] \bigr) \biggr)  +  \bar{c}(x', x) - \bar{c}(x', x)^2  \biggr]  f(x')  ,
        \label{eq:var_hat_f}
    \end{multline}
    where $\bar{c}(x', x) = \sum_{j \in [k]} \frac{c_j(x', x)}{k}$.
    \label{lem:var_hat_f}
\end{lemma}

Combining Lemma \ref{lem:var_hat_f} and Property \ref{pro:ldp_3},  we can derive:

\begin{theorem}
    If $R$ satisfies Properties \ref{pro:ldp_1} - \ref{pro:ldp_3}, we have
    \begin{multline}
        Var( \hat{f}(x) ) = \frac{m^2}{(m - 1)^2 n}  \biggl[ \sum_{x' \in [d] \backslash x} f(x') \\ \biggl( \bar{c}(x', x) Var(R|=) + (1 - \bar{c}(x', x)) Var(R|\ne)  + \\ \bar{c}(x', x)(1 - \bar{c}(x', x)) \biggr)   + f(x) Var(R|=) \biggr] ,
        \label{eq:var_hat_f_sym}
    \end{multline}
    where $Var(R | =)$ and $Var(R | \ne)$ are defined in Property \ref{pro:ldp_3}.
\end{theorem}

 If the hashing family $\mathcal{H}$ is pairwise independent, we further have:

\begin{corollary}
    If $R$ satisfies Properties \ref{pro:ldp_1} - \ref{pro:ldp_3} and $\mathcal{H}$ is pairwise independent (i.e., $\forall x, x': \bar{c}(x', x) = \frac{1}{m}$), we have
    \begin{multline}
        Var( \hat{f}(x) ) 
          \\ = \frac{m}{(m - 1)^2 n}  \biggl[ \bigl( 1 - f(x) \big) \biggl(  Var(R|=) + (m - 1) \\ Var(R|\ne)  + \frac{m - 1}{m} \biggr)   + m f(x) Var(R|=) \biggr] .
        \label{eq:var_hat_f_pairwise}
    \end{multline}
\end{corollary}

The lemma, theorem, and corollary of this section are proven in Appendix \ref{sec:var_proof}.

\subsection{Accuracy and Precision Analysis}
\label{sec:precision_analysis}

The mean squared error (MSE) measures the accuracy and precision of $\hat{f}(x)$, which is defined as
\begin{align}
    MSE(\hat{f}(x)) & = E[(\hat{f}(x) - f(x))^2]  \notag \\  
    & = E[ \hat{f}(x) - E[\hat{f}(x)] ]^2 +  ( E[\hat{f}(x)] - f(x) )^2 \notag
    \\ & =  Var(\hat{f}(x)) + ( E[\hat{f}(x)] - f(x) )^2 \label{eq:mse}
\end{align}

When $\mathcal{H}$ is pairwise independent, we have $E[\hat{f}(x)] - f(x) = 0$, so $MSE(\hat{f}(x)) = Var(\hat{f}(x))$. If a hashing family is not pairwise independent but still satisfies $\forall (x, x') \in [d]^2: \bar{c}(x, x')$ is identical, we can create an unbiased frequency estimator of $f(x)$ as

\begin{equation}
   g(x)  = \frac{m'}{ n (m' - 1) } \bigl[ \sum_{i \in [n]} \hat{y}^{(i)}[h_{j^{(i)}}(x)] \bigr]  - \frac{1}{m' - 1},
\end{equation}

\noindent where $m' = \frac{1}{\bar{c}(x, x')}$.  $g(x)$ has the following properties:
\begin{theorem}
\begin{equation}
    E[g(x)] = 0
\end{equation}
and
\begin{multline}
    MSE(g(x)) = \frac{1}{n(1 - \bar{c}(x, x'))^2}  \biggl[ (1 - f(x)) \\ \biggl( \bar{c}(x', x) Var(R|=) + (1 - \bar{c}(x', x))  Var(R|\ne) \\  + \bar{c}(x', x)(1 - \bar{c}(x', x)) \biggr)   + f(x) Var(R|=) \biggr]  .
\end{multline}
\label{thm:g_x_def}
\end{theorem}

\noindent The formal proof is similar to that of the original estimator $\hat{f}(x)$, so the proof is omitted here. The rationale is that the original estimator $\hat{f}(x)$ assumes $\bar{c}(x, x')$ being $\frac{1}{m}$, but here we integrate $\bar{c}(x, x')$ into the estimation and replace $\frac{1}{m}$ with $\bar{c}(x, x')$.

Besides MSE, one may also use the confidence interval (CI) of $\hat{f}(x)$ to measure precision. If $\hat{f}(x)$ is unbiased, its CI is the same as its concentration bound. Moreover, $\hat{f}(x)$ is equivalent to the sum of bounded random variables (i.e., $\sum \hat{y}$), so its bounds can be derived from the Chernoff bound. Below is the concentration bound for $\hat{f}(x)$ from the unbiased CMS using randomized response (referred to as CMS+RR):
\begin{theorem}
    CMS+RR with pairwise-independent hashing satisfies $\forall \alpha \in [0, \sqrt{\frac{e^{\epsilon} n}{m - 1}}]$,
    \begin{multline*}
        Pr \bigl[ |\hat{f}(x) - f(x)| \ge \alpha \sqrt{Var(\hat{f}(x))}  \bigr] \\ \le 2 \exp(- \frac{\alpha^2}{3} \cdot \frac{m - 1}{e^{\epsilon} + m - 1}) .
    \end{multline*}
    \label{thm:cms_rr_bound}
\end{theorem}

\noindent This theorem is proven in Appendix \ref{sec:proof_cms_rr_bound}. Notably, this theorem focuses solely on CMS+RR, as Section \ref{sec:optimized_cms} will demonstrate that it outperforms CMS implementations with other known perturbations.

\subsection{Optimizing CMS}
\label{sec:optimized_cms}

This section explores the parameters of CMS to reduce the $l_1$ loss, $l_2$ loss, and the worst-case MSE defined in Section \ref{sec:frequency_estimation}. It is notable that $l_2$ loss is connected to MSE as follows:
\begin{multline}
    l_2(\hat{f}) = \max_{\forall \mathbf{X}: |\mathbf{X}| = n} \sum_{x \in [d]} E[ (\hat{f}(x) - f(x) )^2 ] \\ = \max_{\forall \mathbf{X}: |\mathbf{X}| = n} \sum_{x \in [d]} MSE(\hat{f}(x)) . \label{eq:l2_mse}
\end{multline}

Additionally, given the Cauchy-Schwarz inequality, we have $l_1(\hat{f}) \le \sqrt{d \: l_2(\hat{f})}$. Thus, both losses are connected to MSE. Based on Eq. \eqref{eq:mse} and Corollary \ref{cor:fx_no_bias}, if $\mathcal{H}$ is pairwise independent, MSE is equivalent to $Var(\hat{f}(x))$, which can be calculated using Eq. \eqref{eq:var_hat_f_pairwise}. Assuming that the privacy factor $\epsilon$ and the dictionary $[d]$ are given, the undetermined variables in this equation are only the LDP process $R$ and the range of a hash function $m$.

We first consider randomized response (RR) as the LDP process, whose variance is as follows:
\begin{multline}
    Var( RR(v)[v'] ) = \\ \frac{1}{ ( e^{\epsilon} - 1 )^2 } 
    \begin{cases}
        e^{\epsilon} (m - 1) & \text{ if } v = v' \\
        e^{\epsilon} + m - 2 & \text{ if } v \ne v' .
    \end{cases} 
    \label{eq:var_rr}
\end{multline}

Now, we optimize $m$ for CMS+RR, starting with the worst-case MSE. Here, we assume that an analyst has prior knowledge of the upper bound of $f(x)$ and derive:

\begin{theorem}
    If $\forall x \in \mathbf{x}: f(x) \le f^*$ is guaranteed to be correct by prior knowledge, where $\mathbf{x}$ are the values of interest, the worst-case MSE among $\mathbf{x}$ for CMS+RR is optimized as
    
    \begin{equation}
        \widehat{MSE}(\hat{f}) = 
        \begin{cases}
            \frac{2(\Delta_{MSE} + e^{\epsilon})}{n(e^{\epsilon} - 1)^2} & \text{ if } f^* \le \frac{1}{2} \\
            \frac{e^{\epsilon/2}}{n(e^{\epsilon/2} - 1)^2} & \text{ if } \frac{1}{2} < f^* \le 1 
        \end{cases}
        \notag ,
    \end{equation}
    where 
    \begin{equation}
        \Delta_{MSE} = e^{\epsilon /2} \sqrt{[(1 - f^*) e^{\epsilon} + f^*][f^* e^{\epsilon} + (1 - f^*)]} , \notag
    \end{equation}
    and the hash-function range $m$ is the closest integer to    
    \begin{equation}
        \begin{cases}
            1 + \frac{\Delta_{MSE}}{f^* e^{\epsilon} + 1 - f^*}  & \text{ if } f^* \le \frac{1}{2} \\
            1 + e^{\epsilon / 2}  & \text{ if } \frac{1}{2} < f^* \le 1 .
        \end{cases}
    \end{equation}
    \label{thm:optimized_mse}
\end{theorem}

\noindent The theorem is proven in Appendix \ref{sec:proof_optimized_cms_rr}. Notably, $f^*$ is prior knowledge of an analyst and could be larger than $\max_x f(x)$, which is unknown to the analyst. When $f^* \ge \frac{1}{2}$, it has the same worst-case MSE as $f^* = 1$, which is guaranteed to be correct. On the other hand, if $f^* < \frac{1}{2}$ and the prior knowledge happens to be incorrect, i.e., $\max_x f(x) > f^*$, we demonstrate that 

\begin{theorem}
    Given the optimized CMS+RR in Theorem \ref{thm:optimized_mse}, if $\delta = \max_x f(x) - f^* >  0$ and $f^* < \frac{1}{2}$, 
    \begin{multline}
        \widehat{MSE}(\hat{f})  = \widehat{MSE}(\hat{f} | f^*) + \frac{\delta (1 - 2 f^*) e^{\epsilon}}{n\Delta_{MSE}} \\ \le \widehat{MSE}(\hat{f} | f^*) + \frac{\delta(1 - 2 f^*)}{n}.
        \label{eq:inaccurate_f_star}
    \end{multline}

    \noindent where $\widehat{MSE}(\hat{f})$ and $\widehat{MSE}(\hat{f} | f^*)$ are the actual worst-case MSE and the worst-case MSE assuming $\max_x f(x) \le f^*$, respectively.
    \label{thm:inaccurate_f_star}
\end{theorem}

\noindent The theorem is proven in Appendix \ref{sec:proof_optimized_cms_rr}. Since $\widehat{MSE}(\hat{f} | f^*) \ge \widehat{MSE}(\hat{f} | f^* = 0) = \frac{4 e^{\epsilon}}{n(e^{\epsilon} - 1)^2}$, as long as $\delta(1 - 2 f^*)= O(\frac{4 e^{\epsilon}}{n(e^{\epsilon} - 1)^2})$, we still have $\widehat{MSE}(\hat{f}) = O(\widehat{MSE}(\hat{f} | f^*))$. Therefore, if the prior knowledge $f^*$ is guaranteed to be correct or its error is bounded by the above condition, considering the prior knowledge  $f^*$ in the $\widehat{MSE}$ calculation is recommended. Section \ref{sec:norm} provides an example of utilizing $f^*$. Otherwise, one can always assume $f^* = 1$ for simplicity. Hereinafter, we assume no prior knowledge of $f(x)$ unless explicitly mentioned, i.e., $f^* = 1$ by default. 

Next, we optimize the $l_1$ and $l_2$ losses:

\begin{theorem}
    The $l_2$ loss and the upper bound of $l_1$ loss of CMS+RR are optimized as
    \begin{equation}
        l_2^*(\hat{f}) = \frac{2 (\Delta_l + d e^{\epsilon})}{n (e^{\epsilon} - 1)^2} \notag ,
    \end{equation}
    and 
    \begin{equation}
        l_1^*(\hat{f}) \le \sqrt{\frac{2d(\Delta_l + d e^{\epsilon})}{n (e^{\epsilon} - 1)^2}} \notag ,
    \end{equation}
    respectively, where 
    \begin{equation}
        \Delta_l = e^{\epsilon / 2} \sqrt{(e^{\epsilon} + d - 1) ( d e^{\epsilon} - e^{\epsilon} + 1)}  , \notag
    \end{equation}
    and $m$ is the closest integer to $1 + \frac{\Delta_l}{e^{\epsilon} + d - 1}$.
    \label{thm:optimized_l1l2}
\end{theorem}

\begin{corollary}
    When $d \gg e^{\epsilon}$, the optimized $l_2$ loss and the upper bound of the optimized $l_1$ loss of CMS+RR approach
    \begin{equation}
        l_2^*(\hat{f}) \rightarrow \frac{4 d e^{\epsilon}}{n (e^{\epsilon} - 1)^2}  \notag ,
    \end{equation}
    and 
    \begin{equation}
        \lceil l_1 \rceil^* (\hat{f}) \rightarrow \frac{2d e^{\epsilon / 2}}{\sqrt{n} (e^{\epsilon} - 1)} \notag ,
    \end{equation}
    respectively, where $\lceil l_1 \rceil$ represents the upper bound of the $l_1$ loss, and the corresponding $m \rightarrow 1 + e^{\epsilon}$.
    \label{cor:optimized_l1l2_large_d}
\end{corollary}

\noindent The theorem and corollary above are proven in Appendix \ref{sec:proof_optimized_cms_rr}. One caveat is that the CMS+RR optimized for $l_1$ / $l_2$ losses is not optimized for worst-case MSE. Assuming $d$ is large enough, and $m = 1 + e^{\epsilon}$, we have
\begin{equation}
    \widehat{MSE}(\hat{f}) = Var(\hat{f}(x)_{RR} | f(x) = 1) = \frac{(e^{\epsilon} + 1)^2}{n(e^{\epsilon} - 1)^2}, \notag
\end{equation}

\noindent which is always larger than the optimized $\widehat{MSE}$ of CMS+RR, i.e., $\frac{e^{\epsilon / 2}}{n(e^{\epsilon} - 1)^2}$. What is worse, even when $\epsilon$ is very large, it only converges to $\frac{1}{n}$, not zero. Therefore, before an analyst estimates the frequencies of a dataset, they need to determine whether to focus on (1) the overall precision (i.e., $l_1$ / $l_2$ losses) or (2) the worst-case precision of the individual estimations (i.e., $\widehat{MSE}$).

Since optimizing $\widehat{MSE}$ and $l_1$ / $l_2$ losses requires different $m$, we will hereinafter refer to the CMS+RR optimized for $\widehat{MSE}$ and $l_1$ / $l_2$ losses as MSE-OCMS+RR and $l$-OCMS+RR, respectively. Additionally, OCMS+RR will be compared with the existing algorithms in Section \ref{sec:compare}.

Now, we evaluate the CMS using other existing LDP processes. A study \cite{wang2017localhashing} summarized the latest frequency estimation algorithms at that time. When considering different privacy regimes, they found that randomized response (RR), symmetric RAPPOR (RP), and asymmetric RAPPOR (a-RP) achieved state-of-the-art precision in some privacy regimes. Though more advanced algorithms exist, they are either as precise as these algorithms in their respective regimes, or they can be converted to CMS+RR (as shown in Section \ref{sec:connection_with_existing}). Thus, in addition to RR, we will also evaluate CMS using RP and a-RP as the LDP process. Since the precision of these two mechanisms is independent of their input cardinality $m$, and $Var(\hat{f}(x))$ decreases with $m$ when $R$ is independent of $m$, the loss functions of CMS with RAPPOR mechanisms are optimized when $m \rightarrow \infty$.  Appendix \ref{sec:proof_optimized_cms_rappor} derives their values as follows:
\newline
{

\centering
\begin{tabular}{| c | c | c | c |}

\hline
 CMS+ & RR & $sRP$ & $aRP$ \\
 \hline
 \makecell{$\widehat{MSE}(\hat{f}|$ \\  $f^* < \frac{1}{2})$}  & $\frac{2(\Delta_{MSE} + e^{\epsilon})}{n(e^{\epsilon} - 1)^2} $ & \cellcolor{red!50} $ \frac{e^{\epsilon/2}}{n(e^{\epsilon/2} - 1)^2}$ & \cellcolor{red!50} $ \frac{f^*(e^{\epsilon}-1)^2 + 4^{\epsilon}}{n(e^{\epsilon} - 1)^2}$ \\
  \hline
 \makecell{$\widehat{MSE}(\hat{f}|$ \\  $f^* \ge \frac{1}{2})$} & $\frac{e^{\epsilon/2}}{n(e^{\epsilon/2} - 1)^2} $ & $ \frac{e^{\epsilon/2}}{n(e^{\epsilon/2} - 1)^2}$ & \cellcolor{red!50} $ \frac{f^*(e^{\epsilon}-1)^2 + 4^{\epsilon}}{n(e^{\epsilon} - 1)^2}$ \\
 \hline
  $l_2$ & $\frac{4 d e^{\epsilon}}{n (e^{\epsilon} - 1)^2}$ & \cellcolor{red!50} $ \frac{d e^{\epsilon/2}}{n(e^{\epsilon/2} - 1)^2}$  & $\frac{4 d e^{\epsilon}}{n (e^{\epsilon} - 1)^2}$ \\
\hline

\hline
\end{tabular}
\newline
}

CMS using RAPPOR mechanisms are less preferred for the following reasons: (1) their performance in reducing worst-case MSE or $l_1$ / $l_2$ losses is outperformed by CMS+RR in some conditions (highlighted in red in the table), and (2) they have very high communication costs (see  Appendix \ref{sec:proof_optimized_cms_rappor} for details).

\subsection{Unbiased CMS with Imperfect Hashing}
\label{sec:imperfect_hashing}

OCMS+RR is built upon an unbiased CMS, which requires its hashing to be pairwise independent. However, the construction of a $K$-wise-independent hashing family requires the range of the hashed values (i.e., $[m]$) to comprise a finite field (Construction 3.32 of \cite{vadhan2012pseudorandomness}). A finite field must have $p^l$ elements \cite{finite_field_wiki}, where $p$ is a prime number and $l \in \mathbb{N}^+$. For instance, if a CMS has $m = 6$, there is no way to construct a pairwise independent hashing family for it. In practice, we can still construct an approximately pairwise-independent hashing family for any $m$. Consider a perfect pairwise independent hashing family with $p^l > d$:

\begin{equation}
    \mathcal{H}_{pi} = \{ h(x) = a_0 + a_1 x \mid a_0, a_1 \in [p^l] \}.
    \label{eq:hash_family}
\end{equation}

\noindent where $a_0$, $a_1$, and $x$ are elements in a finite field $\mathbb{F}(p^l)$, so the addition and multiplication in Eq. \eqref{eq:hash_family} are all field operations. The range of any $h(x)$ is also the field $\mathbb{F}(p^l)$. One approach to construct $p^l$ is to find a prime number $p' > d$ and set $l = 1$, i.e., $p^l = p'$. In this case, the field operations of $\mathbb{F}(p')$ simply require every operation to modulo $p'$ \cite{menezes2018handbook}. However, finding a prime number when $d$ is very large can be time-consuming. On the other hand, we can find any $l$ such that $2^l > d$, and set $p = 2$, and the field operations of $\mathbb{F}(2^l)$ are introduced in \cite{menezes2018handbook}.

Subsequently, we construct a new hashing family based on $\mathcal{H}_{pi}$:
\begin{equation}
    \mathcal{H}_{api} = \{ h'(x) = h(x) \bmod m \mid h(x) \in \mathcal{H}_{pi} \}. \notag
\end{equation}

\noindent where $api$ represents approximately pairwise independent. Notably, the finite field size $p^l$ should also be no less than $m$. Since $\forall h \in \mathcal{H}_{pi}$, $\forall y \in [m]$: $Pr(h(x) = y) = p^{-l}$, and assuming $p^l = q m + r$, where $q \in \mathbb{N}$ and $r \in \mathbb{N}$, we will have $r$ hashed values with $Pr(h'(x) = y) = \frac{q + 1}{p^l}$ and the remaining $m - r$ hashed values with $Pr(h'(x) = y) = \frac{q}{p^l}$. Thus, we have:

\begin{lemma}
    $\forall x_1, x_2 \in [d]^2$:
\begin{multline}
    Pr_{h' \in \mathcal{H}_{api}}(h'(x_1) = h'(x_2)) \\ = r (\frac{q + 1}{p^l})^2 + (m - r) (\frac{q}{p^l})^2 =  \frac{(2q + 1)r + mq^2}{(m q + r)^2} , \notag
\end{multline}
    \noindent where $h' \in \mathcal{H}_{api}$ represents uniform sampling.
\end{lemma}

Notice that $Pr_{h' \in \mathcal{H}_{api}}(h'(x_1) = h'(x_2)) = \bar{c}(x_1, x_2)$, which is a constant and can be substituted into the $g(x)$ in Theorem \ref{thm:g_x_def}, where $m' = \frac{(m q + r)^2}{(2q + 1)r + mq^2}$. As a result, $g(x)$ is an unbiased estimator of $f(x)$. We can further prove that

\begin{theorem}
    When 
    \begin{equation}
    \frac{2m Var(R|=) + m - 1}{n(m-1)^3} \frac{m}{ \tau Var( \hat{f}(x) ) } \le  (2q + 1)^2 , \notag
\end{equation}

we have 
\begin{equation}
    Var(g(x | \mathcal{H}_{api})) < (1 + \tau) Var(\hat{f}(x)) , \notag
\end{equation}
where $g(x | \mathcal{H}_{api})$ and $\hat{f}(x)$ represent the $g(x)$ with $\mathcal{H}_{api}$ and a standard estimator of $f(x)$ with a pairwise-independent hashing family.
\label{thm:var_g_x_approach_f_x}
\end{theorem}

Applying the theorem to OCMS+RR, we can derive:

\begin{corollary}
\label{cor:g_close_to_f}
    When $2q + 1 > \sqrt{\frac{1}{\tau}}$, we have
    \begin{align}
        MSE(g) & < (1 + \tau) \widehat{MSE}(\hat{f}) \notag \\
        l_2(g) & < (1 + \tau) l_2^*(\hat{f}) \notag
    \end{align} 
    for MSE-OCMS+RR and $l$-OCMS+RR, respectively.
\end{corollary}

The theorem and corollary are proven in Appendix \ref{sec:proof_imperfect_hashing}. Notably, the condition of this corollary is independent of $\epsilon$. If $\tau = 0.01$, we have $q > 4.5$ to satisfy the two corollaries above. Therefore, one could use $g(x)$ to approximate $\hat{f}(x)$ if $p^l \ge 5 m$. Since $a_0$ and $a_1$ of $\mathcal{H}_{api}$ can be any numbers in $\mathbb{F}(p^l)$, we have $|\mathcal{H}_{api}| = (p^l)^2 = p^{2 \lceil \log_p \max \{d + 1, 5 m \}  \rceil}$, and $2 \lceil \log_2 \max \{d + 1, 5 m \} \rceil$ bits are sufficient to represent a hash function in $\mathcal{H}_{api}$ if $p = 2$. 

\subsection{Implementation}
\label{sec:implementation}

\begin{algorithm}
    \caption{OCMS+RR (Part 1)}
    \label{algo:ocms}
  \begin{algorithmic}[1]
    \State $d$: dictionary size
    \State $\epsilon$: privacy factor
    \State Mode := either MSE-OCMS+RR or $l$-OCMS+RR
    \State $f^*$: prior knowledge on the upper bound of $f(x)$. Default value is one.
    \State $\mathit{PRIME} = 2^{64} - 59$ // Largest prime in uint64
    \State
    \Function{FiniteFieldSize}{$d$, $m$}
        \State $v \coloneq \max\{d + 1, 5 m \}$ // see Corollary $\ref{cor:g_close_to_f}$ for why $5m$
        \If{$v \le \mathit{PRIME}$}
            \State\Return $\mathit{PRIME}$
        \EndIf
        \State\Return $2^{\lceil \log_2(v)\rceil}$
    \EndFunction
    \State
    \Function{HashRange}{$\epsilon, d, f^*$, Mode}
        \If{Mode is MSE-OCMS+RR}
            \If{$f^* \ge \frac{1}{2}$}
                \State\Return round( $e^{\epsilon / 2} + 1$ )
            \EndIf
            \State $\Delta \coloneq  e^{\epsilon /2} \sqrt{[(1 - f^*) e^{\epsilon} + f^*][f^* e^{\epsilon} + (1 - f^*)]}$
            \State\Return round(  $ \frac{\Delta}{f^* e^{\epsilon} + 1 - f^*} + 1$ )
        \ElsIf{Mode is $l$-OCMS+RR} 
            \State $\Delta \coloneq e^{\epsilon / 2} \sqrt{(e^{\epsilon} + d - 1) ( d e^{\epsilon} - e^{\epsilon} + 1)}  $
            \State\Return round( $ \frac{\Delta}{e^{\epsilon} + d - 1} + 1$ )
        \EndIf
    \EndFunction
    \State
    \State $m \coloneq $ \Call{HashRange}{$\epsilon, d, f^*$, Mode}
    \State $p_\mathcal{H} \coloneq $ \Call{FiniteFieldSize}{$d, m$} // i.e., $p^l$ in Section \ref{sec:imperfect_hashing}.
    \State
    \Function{Hash}{$a_0, a_1, x, m, p_\mathcal{H}$}
        \If{$p_\mathcal{H} = \mathit{PRIME}$}
            \State $v \coloneq (a_0 + a_1 x) \mod{ p_\mathcal{H} } $
        \Else
            \State $\mathit{op} \coloneq \text{FiniteFieldOperation}( p_\mathcal{H} )$
            \State $v := \mathit{op.add}(a_0, \mathit{op.mul}(a_1, x))$
        \EndIf
        \State\Return $v \mod{ m }$
    \EndFunction
    \State
    \State // return $Q[v, u]$ in Section \ref{sec:ldp}.
    \Function{Decode}{$u$: perturbed value, $v$: value of interest, $m$, $\epsilon$}
        \If{$u = v$}
            \State\Return $\frac{e^{\epsilon} + m - 1}{e^{\epsilon} - 1}$
        \EndIf
        \State\Return $\frac{-1}{e^{\epsilon} - 1}$
    \EndFunction
    \State
    \Function{Client}{$x$: true value of a client, $X^{(i)}$}
        \State Uniformly Sample $a_0$ and $a_1$ from $[0, p_\mathcal{H} - 1]$.
        \State $ y \coloneq $ \Call{Hash}{$a_0, a_1, x, m, p_\mathcal{H}$} // i.e., $h_{j^{(i)}}(X^{(i)})$
        \State \Return \Call{RandomizedResponse}{$y$, $m$, $\epsilon$}, $a_0, a_1$
    \EndFunction
    \algstore{ocms}
  \end{algorithmic}
\end{algorithm}

\begin{algorithm}
    \ContinuedFloat
    \caption{OCMS+RR (Part 2)}
  \begin{algorithmic}[1]
    \algrestore{ocms}
    \Function{Server}{$x$: value of interest, $\mathbf{C}$: all responses from the clients}
        \State $\hat{y}_{tot} \coloneq 0$
        \For{$z, a_0, a_1$ in $\mathbf{C}$}
            \State $ h_x \coloneq $ \Call{Hash}{$a_0, a_1, x, m, p_\mathcal{H}$} // i.e., $h_{j^{(i)}}(x)$
            \State $\hat{y} \coloneq$ \Call{Decode}{$z$, $h_x$, $m$, $\epsilon$} // i.e., $\hat{y}^{(i)}[h_{j^{(i)}}(x)]$
            \State $\hat{y}_{tot} $ +=  $ \hat{y} $
        \EndFor
        \State $n \coloneq$ \Call{Length}{$\mathbf{C}$}
        \State $q , \; r \coloneq \lfloor \: p_\mathcal{H} / m \: \rfloor, \;\; p_\mathcal{H} \mod{m} $
        \State $m' \coloneq \frac{(m q + r)^2}{(2q + 1)r + mq^2}$
        \State\Return $ \frac{m'}{ n (m' - 1) } \hat{y}_{tot}  - \frac{1}{m' - 1} $
    \EndFunction
  \end{algorithmic}
\end{algorithm}

OCMS+RR (Algorithm \ref{algo:ocms}) is implemented to not only adapt the optimized hash range $m$ proposed in Section \ref{sec:optimized_cms}, but also the approximately pairwise independent hashing family $\mathcal{H}_{api}$ and the corresponding estimator $g(x)$ in Section \ref{sec:imperfect_hashing}.  The \textproc{Client} function perturbs each object $X^{(i)}$, and their perturbed values, along with their hashing functions, are sent to the server. The server then calls \textproc{Server} to estimate the frequency of any $x$, i.e., $f(x)$. In particular, when the dictionary size $d$ is small, i.e., $d < 2^{64} - 59$, the algorithm constructs the hashing family with $p^l$ as a prime number. Otherwise, it computes the minimum $2^l > d$ as $p^l$.  The source code of OCMS+RR is available at \url{https://github.com/mingen-pan/CountMeanSketch}).

\subsection{Issues in the original CMS}
\label{sec:correct_original_cms}

Most frequency estimators ensure that the estimated frequency converges to the actual frequency as the dataset size increases, provided each object is queried once. However, this section demonstrates that the estimation from the original CMS exhibits a constant bias, even with an infinitely large dataset.

The original CMS randomly generates $k$ hash functions, and each hash function $h_j(x)$ is generated by sampling $d$ mutually independent random variables, each uniformly distributed among $[m]$. Subsequently, $\forall x, x' \in [d]$ with $x \ne x'$, $c_j(x, x')$ can be considered as a Bernoulli random variable with a probability of $\frac{1}{m}$ of being one. Since the $k$ hash functions are generated independently, $\sum_{j \in [k]} \frac{c_j(x, x')}{k}$ is equivalent to a binomial random variable, whose expectation is $\frac{1}{m}$. However, the probability of the output of a binomial distribution exactly equaling its expectation is very low, so $\sum_{j \in [k]} \frac{c_j(x, x')}{k}$ will likely deviate from $\frac{1}{m}$ to some extent (see Slud's Inequality \cite{slud1977distribution}). Given Corollary \ref{cor:fx_no_bias}, $E[\hat{f}(x)]$ will likely deviate from $f(x)$ to some extent too, i.e., $\hat{f}(x)$ has bias. 

To evaluate the degree of bias, when considering $\forall c_j(x, x')$ as random variables, we can derive:
\begin{theorem}
    \begin{equation}
        \mathop{E}_{ \forall c_j(x, x') } [ E[\hat{f}(x)] ] = f(x)
    \end{equation}
    and 
    \begin{equation}
        \mathop{Var}_{ \forall c_j(x, x') } ( E[\hat{f}(x)] ) = \frac{1}{(m-1)k} \sum_{x' \in [d] \backslash x} f(x')^2 .
        \label{eq:var_e_hat_f_x}
    \end{equation}

    \label{thm:exp_and_var_of_bias}
\end{theorem}

\noindent The proof is presented in  Appendix \ref{sec:proof_exp_and_var_of_bias}. The variance indicates that the bias is $O(\frac{1}{\sqrt{(m - 1) k}})$. However, the original CMS paper \cite{apple_privacy} derived $E[\hat{f}(x)]$ to be unbiased. We found that they treated $c_j(x, x')$ as independent random variables for every pair of $X^{(i)}$ and $X^{(i')}$, which is incorrect because all pairs with the same values $x$ and $x'$ will always have the same $c_j(x, x')$.

The original paper also has an issue in its variance calculation, where they required the hashing family $\mathcal{H}$ to be 3-wise-independent because they assumed any $X^{(i_1)}$ and $X^{(i_2)}$ were correlated, which was problematic. In contrast, the non-private Count Sketch \cite{charikar2002finding} only requires pairwise independence, which is consistent with our work. Notably, the LDP perturbation only affects the precision after the hashing, so it should not alter the pairwise-independence requirement for the hashing family.

\section{Connection with Existing Frequency Estimators}
\label{sec:connection_with_existing}
This section demonstrates that various frequency estimators proposed in the previous literature are equivalent or closely connected to CMS. First, we define the Transformation-and-Sampling Framework (TnS Framework) as follows:

\begin{definition}
(TnS Framework) 

1. Transform the original value $x$ to a vector of elements $M[x]$, i.e., a row of a matrix $M$, where the $j$-th element belongs to the finite set $\mathcal{V}_j$.

2. Uniformly sample an element from the vector $M[x]$, denoted as $M[x][j]$.

3. Perturb the sampled element $M[x][j]$ within $\mathcal{V}_j$ using some LDP mechanism.
\label{def:tns}
\end{definition}

The above process is equivalent to (1) uniformly sampling a column from $M$, denoted as $M[:, j]$, (2) mapping the original value $x$ to $M[x, j]$, and (3) perturbing it. The column here serves as a hash function, and all the columns of $M$ comprise a hashing family. Thus, we have proved:

\begin{theorem}
    An algorithm belonging to the TnS Framework is equivalent to CMS .
    \label{thm:tns}
\end{theorem}

\subsection{Hadamard Encoding}
\label{sec:hadamard_encoding}

Hadamard Encoding (HE) is an algorithm that (1) creates a Hadamard matrix with a size larger than the dictionary size $d$, (2) select a row from the Hadamard matrix with an index equal to $x + 1$, and (3) uniformly samples an element from the selected row, and (4) perturbs the element using randomized response. It was originally proposed by \cite{bassily2015hadamard} and has been widely used in the literature, e.g., \cite{apple_privacy, chen2020rhr}.

It is evident that HE belongs to the TnS framework, so it is equivalent to CMS. Additionally, Appendix \ref{sec:decode_hardmard} proves that the decoding process of HE is equivalent to Eq. \eqref{eq:estimator_def}. Moreover, all the rows of a Hadamard matrix, excluding the first element, comprise a 3-wise-independent hashing family \cite{hardmard_lecture}, mapping to either -1 or 1. Therefore, HE is equivalent to the CMS+RR using pairwise-independent hashing with $m = 2$.

\subsection{Hadamard Response}

Hadamard Response was originally proposed by \cite{acharya2019hr} and later developed in \cite{acharya2019public_hr}. It has two equivalent versions, and the version in \cite{acharya2019public_hr} is presented here: (1) publicly and randomly selects a row from a Hadamard matrix with size larger than the dictionary size $d$, (2) asks if the original value is mapped to one in the selected row (1 means yes, 0 means no), and (3) perturbs the answer using randomized response. It is straightforward to see that Step 1 is equivalent to sampling a hash function, and Step 2 is equivalent to hashing the original value. Additionally, its decoding process is equivalent to HE. Therefore, this algorithm is also equivalent to the CMS+RR using pairwise-independent hashing with $m = 2$.

\subsection{Recursive Hadamard Response}
\label{sec:rhr}

Recursive Hadamard Response (RHR) \cite{chen2020rhr} is an improvement of Hadamard Encoding, increasing precision by communicating more bits. It (1) splits the original value $x$ into $x // 2^{b-1}$ (floor division) and $x \bmod 2^{b-1}$; (2) creates a Hadamard matrix with a size larger than $d // 2^{b-1}$, (3) select a row from the Hadamard matrix with an index equal to $(x // 2^{b-1}) + 1$, and (4) uniformly samples an element from the selected row as $s$, and (5) perturbs $s \times (x \bmod 2^{b-1})$ using randomized response.

Steps 2 - 5 can be fitted in the TnS framework, mapping $x$ to $2^b$ possible values. However, the collision probability of two inputs is not uniform. If two inputs $x_1$ and $x_2$ have the same mod (i.e., $x_1 \bmod 2^{b-1} = x_2 \bmod 2^{b-1}$), their collision probability is $1/2$. Otherwise, the probability is zero. In CMS, any pair of inputs have the same collision probability of $1/m$, which is $2^{-b}$ given $2^b$ hashed values. Thus, RHR is a skewed version of CMS+RR.

If the dataset is randomly generated, i.e., every pair of objects may have the same mod with a probability of $2^{b-1}$, then the hash collision probability becomes $2^{-b}$. Note that \cite{chen2020rhr} sets $b = \epsilon$ if the dictionary is large enough. This is equivalent to $m = e^{\epsilon}$ in CMS, which is close but not identical to the optimized $m$ used by OCMS+RR, so the precision of RHR at most has the same order as OCMS+RR.

However, we can always generate a dataset with all values having the same mod when divided by $2^{b-1}$. This makes all these values have a probability of $1/2$ to collide with one another. Therefore, RHR in this case will perform worse than regular Hadamard Encoding, because they have the same collision probability but the randomized response of RHR considers more possible values, leading to higher variance. In the context of CMS, the MSE of RHR is equivalent to setting $\bar{c}(x, x') = \frac{1}{2}$ while setting $R$ as a randomized response with $m > 2$ in Eq. \eqref{eq:var_hat_f_sym}. Section \ref{sec:kosarak} demonstrates that even a real dataset will encounter this problem if it contains values with high frequency.

\subsection{Local Hashing}

Local Hashing \cite{wang2017localhashing} asks each client to (1) publicly and randomly generate a hashing function mapping $[d]$ to $[m]$, (2) hash the original value, and (3) perturb the hashed value using randomized response. It is straightforward to see that randomly generating a hashing function is equivalent to randomly selecting a hash function from the hashing family in CMS. Also, its decoding process is identical to Eq. \eqref{eq:estimator_def}. Therefore, this algorithm is equivalent to CMS. The original work \cite{wang2017localhashing} assumed $f(x) \approx 0$ when optimizing the MSE of their algorithm, but Section \ref{sec:optimized_cms} demonstrates that both $f(x) = 0$ and $f(x) = 1$ could be the maximum points of the MSE. Moreover, they did not optimize the communication cost of sending a hash function to the server, which will be discussed in Section \ref{sec:compare}.

\subsection{Recursive CMS}
\label{sec:recursive_cms}
When the private CMS was initially proposed in \cite{apple_privacy}, it suggested an algorithm to (1) map an original value to $[m]$ where $m$ is large, and then (2) use Hadamard Encoding to perturb the hashed value. Section \ref{sec:hadamard_encoding} has demonstrated that Hadamard Encoding is equivalent to CMS, so the CMS with Hadamard Encoding (CMS+HE) forms a recursive CMS. Generally speaking, the perturbation algorithm $R$ in a CMS can also be another CMS, and we will prove that a recursive CMS is equivalent to a regular CMS.

Here, we assume all the CMS involved are unbiased, i.e., the hashing family is pairwise independent. Suppose the first CMS hashes the original values to $[m_1]$, and its perturbation algorithm is another CMS hashing its input to $[m_2]$. Given pairwise independence, the collision probability $\bar{c}(x, x')$ of two distinct input values is $\frac{m_1 + m_2 - 1}{m_1 m_2}$. Therefore, the recursive CMS is equivalent to a regular CMS with $m = \frac{m_1 m_2}{m_1 + m_2 - 1}$.

Returning to the CMS+HE, it is equivalent to the CMS+RR with $m' = \frac{2m}{m + 1}$, where $m'$ and $m$ are the hashing ranges of the equivalent and original CMS, respectively. If $m$ is large enough, it is equivalent to the CMS+RR with $m' = 2$ (i.e., Hadamard Encoding).

\subsection{RAPPOR}

The encoding of RAPPOR \cite{rappor} is equivalent to CMS if each object in RAPPOR belongs to its own cohort and there is only one hash function in each cohort. The hashing of RAPPOR and CMS can be considered a linear transformation from $\mathcal{R}^d$ (frequencies of original values) to $\mathcal{R}^{km}$ (frequencies of the hashed values). For an unbiased CMS (Corollary \ref{cor:fx_no_bias}), we have $km = d^2m > d$, so CMS can derive the original frequencies without bias. However, RAPPOR utilizes gradient descent to find the original frequencies and does not require $km \ge d$. The original RAPPOR\cite{rappor} even chooses small $km$ on purpose to reduce computation. Unfortunately, this is problematic because multiple values in a high-dimension simplex will map to the same value in a low-dimension simplex, and the value RAPPOR finds may not be the original value.

\section{Discussion}
\subsection{Precision and Communication Cost}
\label{sec:compare}

Tables \ref{table:precision} and \ref{table:communication} compare the precision and communication cost among various frequency estimators, respectively. We will demonstrate that OCMS+RR is the leading algorithm when considering both precision and communication cost.

\begin{table}[ht]
\centering
\begin{threeparttable}
\begin{tabular}{| c | c | c | c|}
\hline
 & $l_1$ & $l_2$ & $\widehat{MSE}$\\
\hline
HE & $\frac{d(e^{\epsilon} + 1)}{e^{\epsilon} - 1}$ & $\frac{d(e^{\epsilon} + 1)^2}{(e^{\epsilon} - 1)^2}$ & $(\frac{e^{\epsilon} + 1}{e^{\epsilon} - 1})^2$ \\
RHR & \cellcolor{green!25}  $\Theta(\frac{d}{\sqrt{\min \{\epsilon, \epsilon^2 \}}})$ & \cellcolor{green!25}  $\Theta(\frac{d}{\min \{\epsilon, \epsilon^2 \}})$ & $\Omega((\frac{e^{\epsilon} + 1}{e^{\epsilon} - 1})^2)$ \\
OLH & \cellcolor{green!50} $\frac{2d e^{\epsilon / 2}}{e^{\epsilon} - 1} $ & \cellcolor{green!50} $\frac{4de^{\epsilon}}{(e^{\epsilon} - 1)^2}$ & $(\frac{e^{\epsilon} + 1}{e^{\epsilon} - 1})^2$ \\
OCMS+RR & \cellcolor{green!50} $\frac{2d e^{\epsilon / 2}}{e^{\epsilon} - 1} $ & \cellcolor{green!50} $\frac{4de^{\epsilon}}{(e^{\epsilon} - 1)^2}$ & \cellcolor{green!50} $\frac{e^{\epsilon/2}}{(e^{\epsilon/2} - 1)^2}$ \\
CMS+HE & $\frac{d(e^{\epsilon} + 1)}{e^{\epsilon} - 1}$ & $\frac{d(e^{\epsilon} + 1)^2}{(e^{\epsilon} - 1)^2}$ & $(\frac{e^{\epsilon} + 1}{e^{\epsilon} - 1})^2$ \\
SS & \cellcolor{green!50} $\frac{2 d e^{\epsilon / 2}}{e^{\epsilon} - 1} $ \hyperref[SS_l1]{*} & \cellcolor{green!50} $\frac{4de^{\epsilon}}{(e^{\epsilon} - 1)^2}$ & $(\frac{e^{\epsilon} + 1}{e^{\epsilon} - 1})^2$  \\
a-RP & \cellcolor{green!50} $\frac{2d e^{\epsilon / 2}}{e^{\epsilon} - 1} $ & \cellcolor{green!50} $\frac{4de^{\epsilon}}{(e^{\epsilon} - 1)^2}$ & $(\frac{e^{\epsilon} + 1}{e^{\epsilon} - 1})^2$ \\
RP & $\frac{d e^{\epsilon/4}}{ e^{\epsilon/2} - 1 }$ & $\frac{d e^{\epsilon/2}}{(e^{\epsilon/2} - 1)^2}$ & \cellcolor{green!50} $\frac{e^{\epsilon/2}}{(e^{\epsilon/2} - 1)^2}$ \\
\hline
\end{tabular}
\caption{Precision comparison among frequency-estimation algorithms when $d \gg e^{\epsilon}$.}
\begin{tablenotes}
    \item Note: HE, RHR, OLH, OCMS+RR, CMS+HE, SS, a-RP, and RP represent Hadamard Encoding, Recursive Hadamard Response, Optimized Local Hashing, Optimized CMS+RR, CMS with Hadamard Encoding, Subset Selection, asymmetric RAPPOR, and symmetric RAPPOR, respectively. Here, OCMS+RR represents MSE-OCMS+RR and $l$-OCMS+RR when comparing $\widehat{MSE}$ and $l_1$ / $l_2$ losses, respectively. The data source is presented in Appendix \ref{sec:data_compare}.
    \item \text{\label{SS_l1}*}: the original value from \cite{ye2018subset} is problematic, and Appendix \ref{sec:data_compare} recalculates the value as shown.
\end{tablenotes}
\label{table:precision}
\end{threeparttable}
\end{table}

\begin{table}[ht]
\centering
\begin{threeparttable}
\begin{tabular}{| c | w{c}{0.6\columnwidth} |}
\hline
 & Communication Cost (bits)\\
\hline
HE &  \cellcolor{green!50}  $\log d$ \\
RHR & \cellcolor{green!50} $\log d + \epsilon$ \\
OLH &  $d \epsilon$ \\
MSE-OCMS+RR & \cellcolor{green!25} $\max \{ 2 \log d + \frac{\epsilon}{2} , \frac{3}{2} \epsilon + 6 \}$ \\
$l$-OCMS+RR & \cellcolor{green!25} $\max \{ 2 \log d + \epsilon , 3 \epsilon + 6 \}$ \\
original CMS & $\Omega(d \log m)$ \\
SS & $\frac{d}{1 + e^{\epsilon}}$ \\
RAPPOR & $d$ \\
\hline
\end{tabular}
\caption{Communication Cost among frequency-estimation algorithms}
\begin{tablenotes}
    \item Note: see Table \ref{table:precision} for the meaning of abbreviations. The communication cost only accounts for the cost between the server and one client. The data source is presented in Appendix \ref{sec:data_compare}.
\end{tablenotes}
\label{table:communication}
\end{threeparttable}
\end{table}

Table \ref{table:precision} shows that the OCMS+RR is leading in reducing $l_1$ loss, $l_2$ loss, and the worst-case MSE, even considering the constant term. Since HE, OLH, and CMS+HE are equivalent to unbiased CMS+RR without the optimized parameters, it is expected that they are not leading in at least one dimension. Only OCMS+RR and RP are leading in the worst-case MSE, but Table \ref{table:communication} shows that RP requires $d$ bits to communicate while OCMS+RR only requires $\max \{ 2 \log d + \frac{\epsilon}{2} , \frac{3}{2} \epsilon + 6 \}$ bits.  If the communication cost is required to be $O(\log d)$, only RHR has the same order of precision as OCMS+RR in $l_1$ and $l_2$ losses. However, empirical study (Section \ref{sec:experiment}) demonstrates that OCMS+RR outperforms RHR in all datasets, consistent with the theoretical analysis in Section \ref{sec:rhr}. Therefore, OCMS+RR is the only leading algorithm for the worst-case MSE, $l_1$ loss, and $l_2$ loss while communicating efficiently.

\subsection{Necessity of Randomness}
\label{sec:necessity_of_randomness}

An algorithm was proposed by \cite{acharya2019hr} to convert the Hadamard Response (a special case of CMS) into a deterministic algorithm, arguing that the randomness could be avoided, thereby decreasing the communication cost. However, we will show that one can always create a counterexample causing the frequency estimation to have a constant bias. Assume that we are interested in the frequency of value $x$, and $f(x)$ happens to be zero (no $x$ in the dataset). Given any hash function (denoted as $h_j$ for convenience), we denote $x_j$ as the value colliding with $x$, i.e., $c_j(x, x_j) = 1$. Subsequently, we set the value of all objects assigned to the $j$-th hash function as $x_j$, and repeat this process for all the hash functions. As a result, we have:

\begin{equation*}
    \sum_{j \in [k]} \sum_{x' \in [d] \backslash x} c_j(X^{(i)}, x) f(x') =  1.
\end{equation*}

\noindent Substituting the above formula and $f(x) = 0$ into Eq. \eqref{eq:expect_fx} results in $E[\hat{f}(x)] = 1$, which proves:

\begin{theorem}
    If the assignment of hash functions to the objects, $\{ h_{j^{(i)}} | i \in n \}$, is constant in CMS, there exists a dataset resulting in
    \begin{equation*}
        E[\hat{f}(x)] - f(x) \ne 0 .
    \end{equation*}
\end{theorem}

The logic behind the theorem is intuitive: CMS utilizes randomness to ensure a stable probability of collision with other values. If one gives up the randomness, an attacker can always generate a dataset to maximize the collision. The reason why \cite{acharya2019hr} believed a deterministic algorithm still worked was that they assumed that each $X^{(i)}$ was sampled from a distribution, so one cannot control the value of $X^{(i)}$. Their algorithm relies on the randomness in distribution sampling instead of the randomness in assigning hash functions.

The next question is how much randomness (in bits) is needed to ensure that CMS will work as expected. Here, we focus only on the randomness in assigning hash functions and assume that the LDP process $R$ always uses its own randomness.

By examining the proof involving hash-function assignments in Appendices \ref{sec:expect_proof} and \ref{sec:var_proof}, we realize $Var(\hat{f}(x))$ is calculated based on $Cov(\hat{y}^{(i_1)}[h_{j^{(i_1)}}(x)], \hat{y}^{(i_2)}[h_{j^{(i_2)}}(x)]) = 0$ for any $i_1$ and $i_2$, which is further derived from the assumption that the hash-function assignments $\mathbf{1}(j^{(i_1)} = j)$ and $\mathbf{1}(j^{(i_2)} = j)$ are pairwise independent. 

Recall that hash-function assignment uniformly samples from $\mathcal{H}$ for each object. This process maps $n$ distinct inputs (object IDs) to $n$ repeatable elements in $[k]$ (recall $k = |\mathcal{H}|$), which can be considered as hashing, denoted as $s$. We require $s$ to be uniformly sampled from a pairwise-independent hashing family $\mathcal{S}$, which guarantees the pairwise independence of $\mathbf{1}(j^{(i_1)} = j)$ and $\mathbf{1}(j^{(i_2)} = j)$ for all $i_1$, $i_2$, and $j$. If we follow Section \ref{sec:imperfect_hashing} to construct a hashing family mapping from $[n]$ to $[k]$, the parameter $p^l$ should exceed $\max\{n, k\}$, which requires $2\lceil \log_2( \max\{n, k\} + 1 )\rceil$ bits to represent one of its mappings. We have proved:

\begin{theorem}
     $2\lceil \log_2( \max\{n, k\} + 1 )\rceil$ random bits are sufficient to generate all the hash-function assignments $[j^{(i)}]_{i \in [n]}$ while ensuring their pairwise independence, which is the prerequisite of Eq. \eqref{eq:var_hat_f}.
    \label{thm:necessity_of_randomness}
\end{theorem}

\subsection{Public or Private Randomness}
\label{sec:public_private_randomness}

Previous literature introduces the concept of public and private randomness as follows: \vspace{4pt}

\noindent \textbf{Public Randomness}: the server generates a string of random bytes, which are accessible by the public (i.e., clients). 

\vspace{2pt} \noindent \textbf{Private Randomness}: the client generates a string of random bytes, which will be used in their algorithm. The client may or may not publish the random bytes.

\vspace{4pt} Previous literature (e.g., \cite{bassily2015hadamard, acharya2019public_hr, chen2020rhr}) had a strong interest in public randomness, because it was believed that public randomness could save communication costs. However, we argue that public and private randomness make no difference for an unbiased CMS, and no communication cost can be saved, for the following reasons: the server and a client (i.e., an object in CMS) have to sync on which hash function is assigned to the client, and they need to either sync (1) the identifier of a hash function, or (2) the assignment generator. There are at least $(d + 1)^2$ hash functions in a pairwise independent $\mathcal{H}_{pi}$, so identifying one of them requires $2\log_2 d$ bits. Meanwhile, Theorem \ref{thm:necessity_of_randomness} shows that at least $2 \log_2 k$ bits are required to represent the assignment generator, where $k$ is the hashing family size, so this is equivalent to requiring at least $4\log_2 d$ bits. Thus, syncing the identifier will be less expensive, and $2\log_2 d$ bits are needed to sync, regardless of which side generates the assignment. Additionally, $\log_2 n$ bits are required if the server and a client decide to sync on the assignment generator, because the client needs to send back the object ID (i.e., $i \in [n]$) to identify itself.

When reviewing the existing literature, the work claiming one-bit communication usually missed the communication cost of syncing the public randomness. For instance, \cite{bassily2015hadamard} proposed a protocol to convert a private-randomness algorithm to a public one. Though the client only sends one bit as a response, it has to sync $b$ bits from the server, where $b$ is the number of bits the private-randomness algorithm would send. Moreover, the server also needs the client/object ID to know where the one bit comes from, which will contribute $\log n$ bits to the communication cost. The same problem also exists in the public-randomness algorithm in \cite{acharya2019public_hr}.

The only scenario where public randomness could save communication costs is when all the clients are clocated, referred to as a client group. The traffic inside a client group is assumed to be free. When utilizing the public randomness, the server will share the hash-function assignment generator with the client group, which costs $\lceil \max\{4\log_2 d, 2 \log_2 n\} \rceil$ bits. Then, the client group shares the generator, calculates their assignment of the hash functions, and replies with their responses in sequence back to the server, which will cost $n \log m \approx n \epsilon$ bits. In this case, the total communication cost between the server and the client group is $\lceil \max\{4\log_2 d, 2 \log_2 n\} \rceil + n \epsilon$. If the client group still uses private randomness, they will return $n (2\log_2 d + \epsilon)$ bits, which is larger than the cost from public randomness. However, the client-group assumption is contradictory to the motivation of LDP, which usually assumes the clients are distributed.

\begin{figure*}[ht!]
\centering
\subfloat[]{\includegraphics[width=0.32\textwidth]{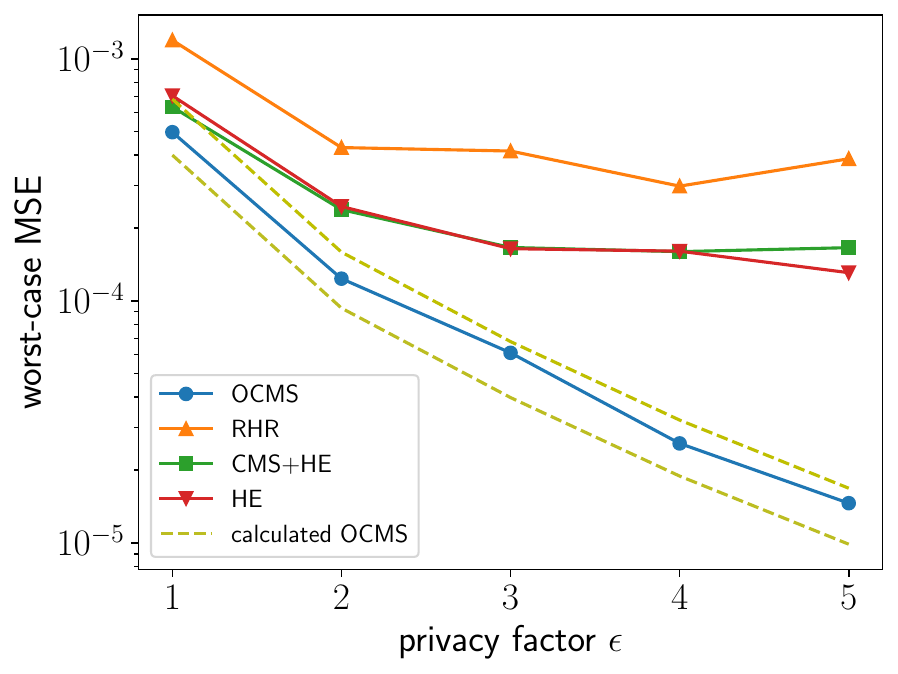}}
\hfill
\subfloat[]{\includegraphics[width=0.32\textwidth]{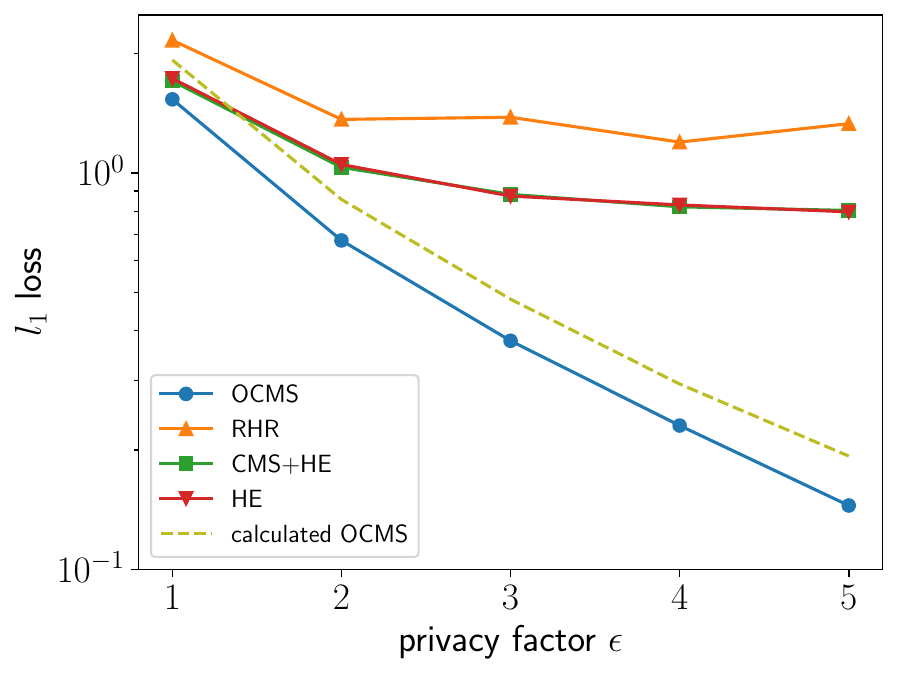}}%
\hfill
\subfloat[]{\includegraphics[width=0.32\textwidth]{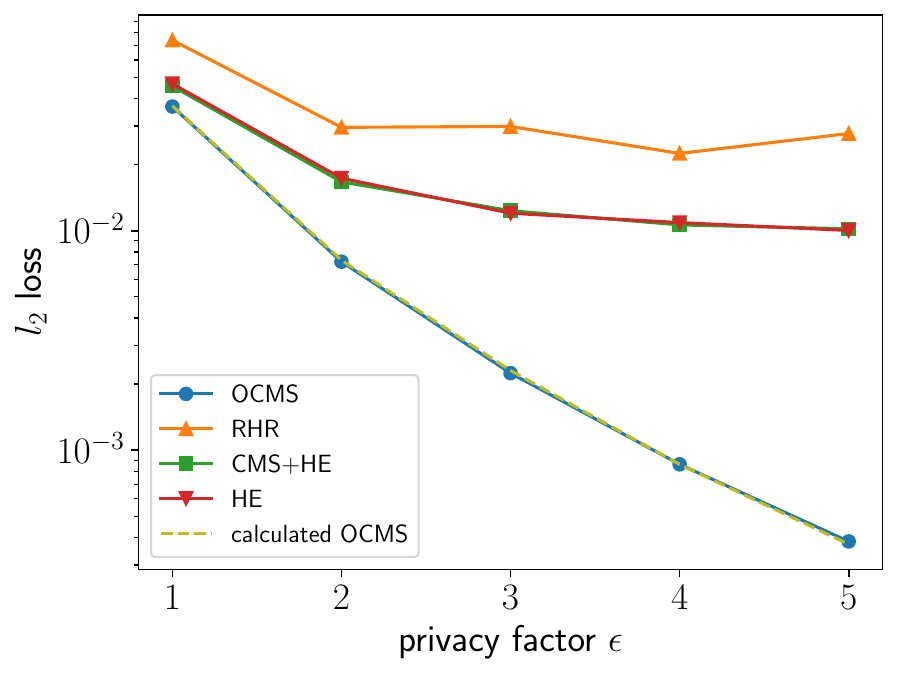}}%

\caption{Worst-case MSE, $l_1$ loss, and $l_2$ loss vs. privacy factor $\epsilon$ given the Zipf dataset. See Section \ref{sec:zipf} for details.}
\label{fig:zipf} 

\subfloat[]{\includegraphics[width=0.32\textwidth]{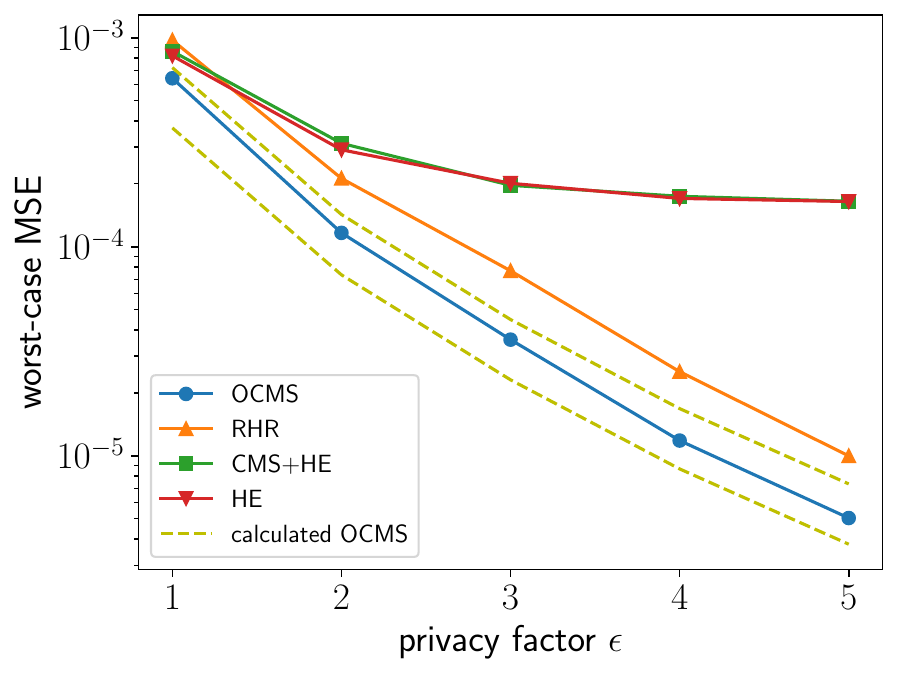}}
\hfill
\subfloat[]{\includegraphics[width=0.32\textwidth]{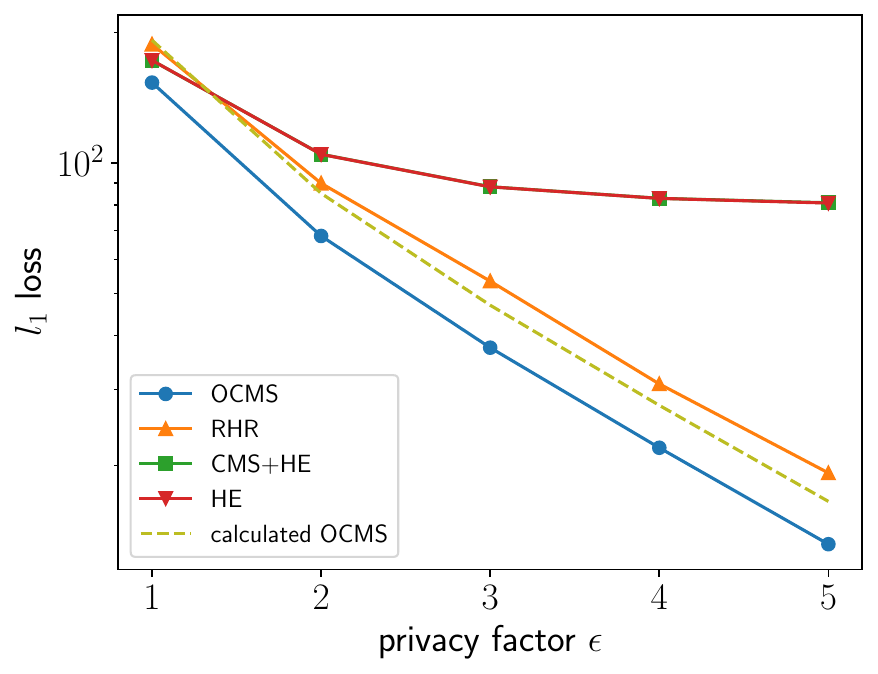}}%
\hfill
\subfloat[]{\includegraphics[width=0.32\textwidth]{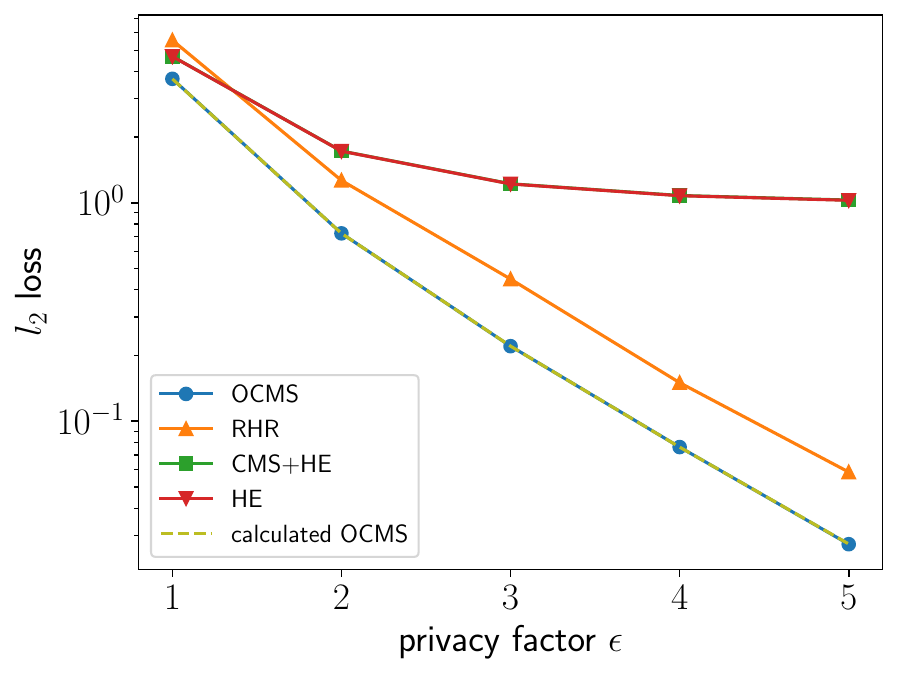}}%

\caption{Worst-case MSE, $l_1$ loss, and $l_2$ loss vs. privacy factor $\epsilon$ given the Gaussian dataset. See Section \ref{sec:norm} for details.}
\label{fig:norm} 

\subfloat[]{\includegraphics[width=0.32\textwidth]{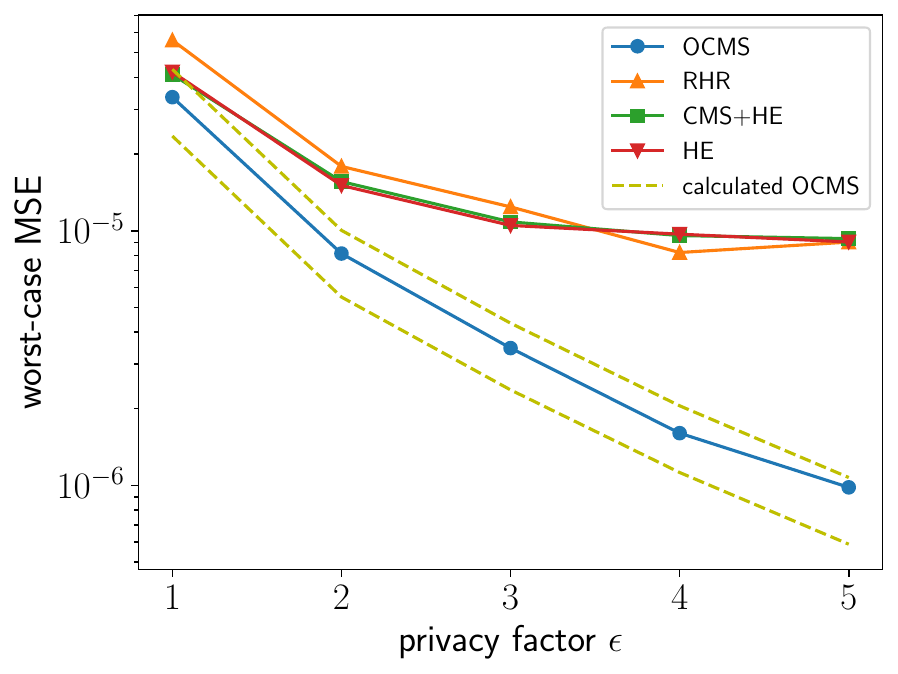}}
\hfill
\subfloat[]{\includegraphics[width=0.32\textwidth]{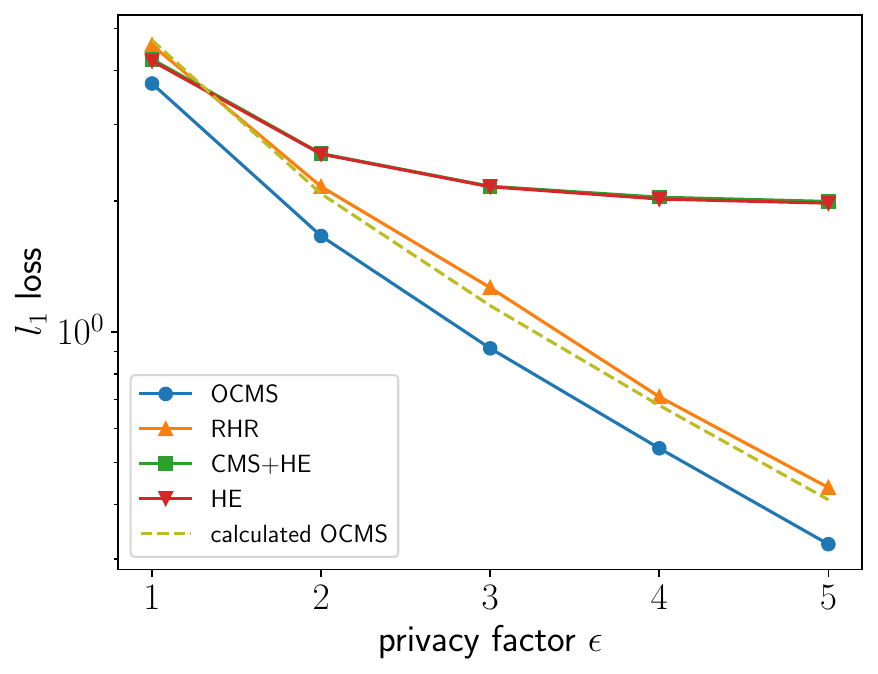}}%
\hfill
\subfloat[]{\includegraphics[width=0.32\textwidth]{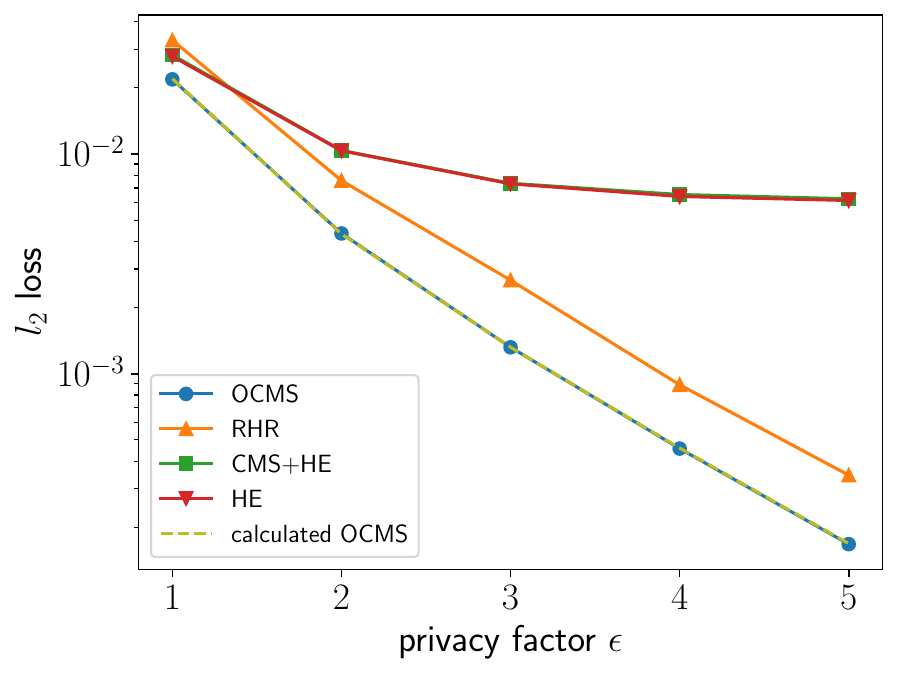}}%

\caption{Worst-case MSE, $l_1$ loss, and $l_2$ loss vs. privacy factor $\epsilon$ given the mini Kosarak dataset. See Section \ref{sec:kosarak} for details.}
\label{fig:kosarak} 
\end{figure*}

\section{Experiment}
\label{sec:experiment}
This section empirically compares the OCMS+RR (implemented following Section \ref{sec:implementation}) with the algorithms listed below, using both synthetic and real-world datasets: \vspace{2pt}

1. Hadamard Encoding (HE): the first practical frequency estimation algorithm.

2. CMS with HE (CMS+HE): the original CMS algorithm with the same parameter as \cite{apple_privacy}.

3. Recursive Hadamard Response (RHR): the state-of-the-art version of the Hadamard Response family. 

\vspace{2pt} Other algorithms listed in Section \ref{sec:compare} are not included due to their very high communication cost, which is impractical for a very large dictionary.

In the following experiments, we use this estimator to estimate $\widehat{MSE}$:
\begin{equation}
    \overline{MSE}(\hat{f}) = \max_{x \in \mathbf{x}} \frac{1}{t} \sum_t (\hat{f}(x) - f(x))^2 ,
\end{equation}

\noindent where $\mathbf{x}$ denotes the values of interest and $t$ denotes the experiment rounds. However, $\overline{MSE}(\hat{f})$ is biased because $\frac{1}{t} \sum_t (\hat{f}(x) - f(x))^2$ of each $x$ is equivalent to a random variable fluctuating around $E[(\hat{f}(x) - f(x))^2]$, and the max operator will choose the largest deviated value. Appendix \ref{sec:upper_bound_mse} provides an upper bound of $\overline{MSE}(\hat{f})$. As long as $\overline{MSE}(\hat{f})$ is between the upper bound and the calculated $\widehat{MSE}(\hat{f})$ (both bounds are plotted), we will consider the experimental $\overline{MSE}(\hat{f})$ consistent with the theoretical value from Theorem \ref{thm:optimized_mse}.

Moreover, we will compute $l_1$ / $l_2$ losses only from a subset of $[d]$, denoted as $\mathbf{x}$. By examining Eq. \eqref{eq:l2_var_general}, one could prove that the $l_1$ / $l_2$ losses from Theorem \ref{thm:optimized_l1l2} are still valid when replacing $d$ with $|\mathbf{x}|$. Note that the calculated $l_1$ loss is an upper bound, so the experimental value is expected to be below it.

All the experiments below will be run 100 times for each privacy factor $\epsilon$ ranging from 1 to 5.

\subsection{Verification of Precision and Equivalence}
\label{sec:zipf}

We use a synthetic dataset sampled from a specific Zipf's distribution, which has a frequency distribution as

\begin{equation*}
    Zipf(r) \propto \frac{1}{r^2} ,
\end{equation*}

\noindent where $r$ is the order of a value ranked by its frequency. The dataset values are chosen deliberately, so all have the same mod for the Recursive Hadamard Response (RHR) (see Section \ref{sec:rhr}). The dictionary size is set to 100k, and 10,000 objects are sampled from the distribution. The $\widehat{MSE}$, $l_1$ loss, and $l_2$ loss are evaluated based on the first 100 most frequent values (assuming they are known in advance). MSE-OCMS+RR with $f^* = 1$ (i.e., no prior knowledge on $f(x)$) and $l$-OCMS+RR are evaluated for $\widehat{MSE}$ and $l_1$ / $l_2$ losses, respectively, while RHR, HE, and CMS+HE are evaluated for all the metrics. Results are plotted in Fig. \ref{fig:zipf}.

MSE-OCMS+RR and $l$-OCMS+RR outperform other algorithms in terms of $\widehat{MSE}$ and $l_1$ / $l_2$ losses, respectively, and their precision aligns with the calculated values. HE and CMS+HE have identical performance in all the metrics, supporting our theory that these two algorithms are equivalent. Consistent with our expectation, RHR performs the worst in all the metrics because we deliberately construct a dataset, though using only public information, to ensure every object has a 50\% probability of colliding with another in RHR, resulting in poorer performance than HE, as analyzed in Section \ref{sec:rhr}.

\subsection{Synthetic Dataset with Prior Knowledge}
\label{sec:norm}

A synthetic database is constructed with 10,000 objects sampled from a normal distribution with a standard deviation of 50. The mean of the distribution is an unknown integer picked from [1000, 9000]. The dictionary size is 10,000, so the values of all the objects are rounded to integers and truncated if they are outside of [0, 10,000]. Through several simulations of this distribution, we are confident that the probability of $\max_x f(x) \le 0.01$ is high. Thus, we set $f^* = 0$ for MSE-OCMS+RR and its calculated $\widehat{MSE}$ is derived from Eq. \eqref{eq:inaccurate_f_star} as:

\begin{equation}
    \widehat{MSE}(\hat{f}) = \frac{1}{n} [\max_x f(x) + \frac{4e^{\epsilon}}{(e^{\epsilon} - 1)^2}] . 
    \notag
\end{equation}

\noindent Meanwhile, $l$-OCMS+RR is evaluated for $l_1$ / $l_2$ losses. RHR, HE, and CMS+HE are also evaluated for $\widehat{MSE}$, $l_1$ loss, and $l_2$ loss. Results are plotted in Fig. \ref{fig:norm}.

Similar to the above, MSE-OCMS+RR and $l$-OCMS+RR outperform other algorithms in terms of $\widehat{MSE}$ and $l_1$ / $l_2$ losses, respectively, and their precision aligns with the calculated values. HE and CMS+HE have identical precision in all the metrics, and they perform the worst. The precision of RHR lies between OCMS+RR and the other algorithms.

\subsection{Real-world Dataset}
\label{sec:kosarak}

Kosarak \cite{kosarak} is a dataset containing click-stream data from a Hungarian online news portal. The dataset is shrunk to 1\% of its original size, denoted as mini Kosarak, which will be used in this section. It contains over 170,000 entries, all of which belong to a dictionary with a size of 26,000. MSE-OCMS+RR with $f^* = 1$ and $l$-OCMS+RR are evaluated for $\widehat{MSE}$ and $l_1$ / $l_2$ losses, respectively, while RHR, HE, and CMS+HE are evaluated for all the metrics. Results are plotted in Fig. \ref{fig:kosarak}.

MSE-OCMS+RR and $l$-OCMS+RR outperform other algorithms in terms of $\widehat{MSE}$ and $l_1$ / $l_2$ losses, respectively, and their precision aligns with the calculated values. HE and CMS+HE have identical precision in all the metrics, and they perform the worst in $l_1$ / $l_2$ losses. RHR has the same order of precision as $l$-OCMS+RR regarding $l_1$ / $l_2$ losses. However, it performs the worst in $\widehat{MSE}$. This is expected because Kosarak has one value whose frequency is above 0.4, which negatively impacts the MSE of other values sharing the same mod with it (see Section \ref{sec:rhr} for a detailed explanation).

\section{Conclusion}

This paper revisits the private CMS algorithm, corrects errors in the expectation and variance calculations in the original CMS paper, and optimizes CMS with randomized response to reduce all kinds of loss functions. The optimized CMS is proven theoretically and empirically to be the leading algorithm for reducing the worst-case MSE, $l_1$ loss, and $l_2$ loss, when requiring the communication cost to be logarithmic to the dictionary size. Moreover, this paper demonstrates that many existing algorithms are equivalent to CMS. Finally, we discuss the necessity of randomness in CMS and show that both public and private randomness have the same communication cost.

\bibliographystyle{IEEEtran}
\bibliography{references}  
%

\appendices

\section{Example on Reconstruction Process}
\label{sec:example_reconstruction}

Suppose a dictionary $\mathbb{V}$ has three possible values: 1, 2, and 3; and another dictionary $\mathbb{U}$ has five possible values: $A, B, C, D$, and $E$. A mechanism $M$ perturbs a variable $V \in \mathbb{V}$ and outputs a value $U \in \mathbb{U}$. The probability that $M$ outputs $U = u$ given $V = v$ is randomly assigned as

\begin{center}
\begin{tabular}{ | c | c | c | c |} 

\hline
\backslashbox{$u$}{$v$} & 1 & 2 & 3  \\
 \hline 
 A & 0.133 & 0.143 & 0.192 \\
 B & 0.031 & 0.212 & 0.658 \\
 C & 0.287 & 0.155 & 0.039  \\
 D & 0.161 & 0.254 & 0.060  \\
 E & 0.389 & 0.236 & 0.051  \\ 
\hline
\end{tabular}
\end{center}

\noindent which is equivalent to matrix $P$. Subsequently, the inverse matrix $Q$ is derived as

\begin{center}
\begin{tabular}{ | c | c | c | c | c | c |} 

\hline
 \backslashbox{$v$}{$u$} & A & B & C & D & E  \\
 \hline 
 1 & 0.366 & -0.088 & 2.289 & -3.629 & 2.267 \\
 2 & 0.010 & -0.393 & -1.928 & 6.625 & -1.294 \\
 3 & 0.332 & 1.560 & 0.476 & -1.996 & 0.240  \\
\hline
\end{tabular}
\end{center}

If $M$ perturbs $V$ and outputs $U = C$, the decoding process will return the column of $Q$ corresponding to $C$, which is $[2.289, -1.928, 0.476]$. If $V = 1$, the probability that $M(V)$ outputs $U = C$ is 0.287. The same probability applies to the event that the reconstruction process $R(V)$, i.e., $Q[: M(V)]$, generates the corresponding column. Denote the column outputted by the reconstruction as $\tilde{V}$. One can verify that the expectation of $\tilde{V}$ equals $[1, 0, 0]$ when $V = 1$, which is consistent with Property \ref{pro:ldp_1}.


\section{Proof regarding Optimizing CMS with RR}
\label{sec:proof_optimized_cms_rr}

We first explore the optimized $m$ for CMS + RR. Substituting Eq. \eqref{eq:var_rr} into Eq. \eqref{eq:var_hat_f_pairwise}, we have
\begin{multline}
    Var(\hat{f}(x)_{RR}) = \frac{1 - f(x)}{n (m - 1)} + \\  \frac{m [(1 - f(x)) (e^{\epsilon} - 1) (2 - m) + m e^{\epsilon}]}{n (m - 1) (e^{\epsilon} - 1)^2} 
    \label{eq:cms_var_rr}
\end{multline}

Let us first study the worst-case MSE of CMS+RR, which is equivalent to the worst-case variance $Var(\hat{f}(x)_{RR})$.  Observing that the variance is linear to $f(x)$, we have $\max_x MSE(\hat{f}(x)) = \max\{Var(\hat{f}(x)_{RR} | f(x) = 0), Var(\hat{f}(x)_{RR} | f(x) = 1) \}$. When $f(x) = 1$, we have

\begin{equation}
    \frac{\partial Var(\hat{f}(x)_{RR} | f(x) = 1)}{\partial f(x)} = \frac{m (m - 2) e^{\epsilon}}{n (m - 1)^2 (e^{\epsilon} - 1)^2} , \notag
\end{equation}

\noindent implying that $Var(\hat{f}(x)_{RR} | f(x) = 1)$ increases monotonically with $m$ when $m \ge 2$. On the other hand, when $f(x) = 0$, we have

\begin{equation}
    \frac{\partial Var(\hat{f}(x)_{RR} | f(x) = 0)}{\partial f(x)} = \frac{(m - e^{\epsilon}- 1) (m + e^{\epsilon} - 1)}{n (m - 1) (e^{\epsilon} - 1)^2} , \notag
\end{equation}

\noindent which indicates a minimum at $m = e^{\epsilon} + 1$ given the range $m \ge 2$. However, when $m = e^{\epsilon} + 1$, $Var(\hat{f}(x)_{RR} | f(x) = 1)$ is always larger than $Var(\hat{f}(x)_{RR} | f(x) = 0)$. Given that $Var(\hat{f}(x)_{RR} | f(x) = 1)$ decreases as $m$ decreases, while $Var(\hat{f}(x)_{RR} | f(x) = 0)$ increases when $m$ decreases within the range $[2, e^{\epsilon} + 1]$, there exists an $m$ making them equal to each other. Solving

\begin{equation}
    Var(\hat{f}(x)_{RR} | f(x) = 0) = Var(\hat{f}(x)_{RR} | f(x) = 1), \notag
\end{equation}

\noindent we have $m = e^{\epsilon / 2} + 1$, and the corresponding variance is

\begin{equation}
    Var(\hat{f}(x)_{RR}) = \frac{e^{\epsilon/2}}{n(e^{\epsilon/2} - 1)^2} . \notag
\end{equation}

\noindent When $ m \in [2, e^{\epsilon / 2} + 1] $, we have $Var(\hat{f}(x)_{RR} | f(x) = 1) < Var(\hat{f}(x)_{RR} | f(x) = 0)$. Also, $Var(\hat{f}(x)_{RR} | f(x) = 0)$ decreases with $m$ in this range. Therefore,when $ m \in [2, e^{\epsilon / 2} + 1] $, we have
\begin{multline*}
    \max_{f(x)} Var(\hat{f}(x)_{RR}) > \max_{f(x)} Var(\hat{f}(x)_{RR} | m  = e^{\epsilon / 2} + 1) \\ = \frac{e^{\epsilon/2}}{n(e^{\epsilon/2} - 1)^2} .
\end{multline*}

\noindent That is, $m$ should never be smaller than $e^{\epsilon / 2} + 1$.

Define $f^*$ as the upper bound of $f(x)$, i.e., $\forall x: f(x) \le f^*$. When $ m \in [e^{\epsilon / 2} + 1, e^{\epsilon} + 1] $, we have proven that $Var(\hat{f}(x)_{RR})$ increases with $f(x)$, so the goal becomes reducing $Var(\hat{f}(x)_{RR} | f(x) = f^*)$. The optimized $m$ can be calculated by

\begin{equation}
  \frac{\partial Var(\hat{f}(x)_{RR} | f(x) = f^*)}{\partial m} = 0  , \notag
\end{equation}

\noindent which results in 
\begin{equation}
    m = 1 + \frac{\Delta_{MSE}}{f^* e^{\epsilon} + 1 - f^*} , 
    \label{eq:best_cms_var_rr_m_small_f}
\end{equation}

\noindent where 
\begin{equation}
    \Delta_{MSE} = e^{\epsilon /2} \sqrt{[(1 - f^*) e^{\epsilon} + f^*][f^* e^{\epsilon} + (1 - f^*)]}  \notag .
\end{equation}

\noindent The corresponding variance is
\begin{equation}
   Var^*(\hat{f}(x)_{RR})  = \frac{2(\Delta_{MSE} + e^{\epsilon})}{n(e^{\epsilon} - 1)^2} \notag
\end{equation}

Notice that Eq. \eqref{eq:best_cms_var_rr_m_small_f} is valid only when $m \in [e^{\epsilon / 2} + 1, e^{\epsilon} + 1]$, which translates to $f^* \le \frac{1}{2}$. When $f^* > \frac{1}{2}$, we have $m < 1 + e^{\epsilon / 2}$, which has been proven to be worse than $m = 1 + e^{\epsilon / 2}$. Therefore, we have

\begin{equation}
    m = 
    \begin{cases}
        1 + \Delta_{MSE} & \text{ if } f^* \le \frac{1}{2} \\
        1 + e^{\epsilon / 2}  & \text{ if } \frac{1}{2} < f^* \le 1 ,
    \end{cases}
    \notag
\end{equation}

\noindent which corresponds to
    
\begin{equation}
    Var^*(\hat{f}(x)_{RR})  = 
    \begin{cases}
        \frac{2(\Delta_{MSE} + e^{\epsilon})}{n(e^{\epsilon} - 1)^2} & \text{ if } f^* \le \frac{1}{2} \\
        \frac{e^{\epsilon/2}}{n(e^{\epsilon/2} - 1)^2} & \text{ if } \frac{1}{2} < f^* \le 1 
    \end{cases}
    \notag .
\end{equation}

Regarding Theorem \ref{thm:inaccurate_f_star}, we observe that $Var(\hat{f}(x)_{RR})$ in Eq. \eqref{eq:cms_var_rr} is linear to $f(x)$, so
\begin{multline*}
     Var(\hat{f}(x)_{RR} | f(x) = f^* + \delta) - Var(\hat{f}(x)_{RR} | f(x) = f^*  ) \\ = \frac{(m - 1)^2 - e^{\epsilon}}{(m - 1) (e^{\epsilon} - 1) n} \delta .
\end{multline*}

\noindent Substituting $m = 1 + \Delta_{MSE}$ into the above equation and simplifying it, we have:
\begin{multline*}
     Var(\hat{f}(x)_{RR} | f(x) = f^* + \delta) - Var(\hat{f}(x)_{RR} | f(x) = f^*  ) \\ = \frac{\delta (1 - 2 f^*) e^{\epsilon}}{n\Delta_{MSE}} .
\end{multline*}

\noindent Since $Var(\hat{f}(x)_{RR})$ increases with $f(x)$ given $m = 1 + \Delta_{MSE}$, we have
\begin{multline*}
    \widehat{MSE}(\hat{f}(x)) = Var(\hat{f}(x)_{RR} | f(x) = f^* + \delta) \\ = \widehat{MSE}(\hat{f} | f^*) +  \frac{\delta (1 - 2 f^*) e^{\epsilon}}{n\Delta_{MSE}} ,
\end{multline*}

\noindent which proves the equality sign of Theorem \ref{thm:inaccurate_f_star}. Given that $\frac{e^{\epsilon}}{\Delta_{MSE}}$ decreases when $f^* \in [0, \frac{1}{2}]$, we have $\frac{e^{\epsilon}}{\Delta_{MSE}} \le 1$, which proves the ``$\le$" sign of   Theorem \ref{thm:inaccurate_f_star}. $\qed$

Now, we will explore the optimized parameters for $l_2$ loss. By considering only the pairwise independence of $\mathcal{H}$, we integrate Eq. \eqref{eq:var_hat_f_pairwise} into Eq. \eqref{eq:l2_mse}:
\begin{multline}
    l_2(\hat{f}) = \sum_{x \in [d]} \frac{m}{n (m - 1)^2} \bigl[  \bigl(  Var(R | =)   + (m - 1) Var(R | \ne) \\ + \frac{m-1}{m} \bigr) (1 - f(x))  + m  Var(R | =) f(x) \bigr] \\
    =  \frac{m}{n (m - 1)^2} \bigl[  \bigl(  Var(R | =)   + (m - 1) Var(R | \ne)  \\ + \frac{m-1}{m} \bigr) (d - 1)  + m  Var(R | =) \bigr]
    \label{eq:l2_var_general}
\end{multline}

Applying RR and Substituting Eq. \eqref{eq:var_rr} into Eq. \eqref{eq:l2_var_general}, we obtain
\begin{equation*}
    l_2(\hat{f}_{RR}) = \frac{ (d + e^{\epsilon} - 1) m^2}{n (m - 1) (e^{\epsilon} - 1)^2}  + \frac{   (d - 1) (2m + e^{\epsilon} - 1)}{n (m - 1) (e^{\epsilon} - 1)} .
\end{equation*}

By studying its derivative with respect to $m$, we can find its minimum when
\begin{equation}
    m = 1 + \frac{\Delta_l}{e^{\epsilon} + d - 1} , \notag
\end{equation}

\noindent where
\begin{equation}
    \Delta_l = \sqrt{(e^{\epsilon} + d - 1) ( d e^{\epsilon} - e^{\epsilon} + 1)} e^{\epsilon / 2} .\notag
\end{equation}

\noindent In practice, we choose the closest integer to $1 + \frac{\Delta_l}{e^{\epsilon} + d - 1}$ as $m$. The corresponding optimized $l_2$ is the one in Theorem \ref{thm:optimized_l1l2}. Given the Cauchy-Schwarz inequality, we have $l_1 \le \sqrt{d \: l_2}$, which derives the optimized $l_1$ in Theorem \ref{thm:optimized_l1l2}. Corollary \ref{cor:optimized_l1l2_large_d} is proven accordingly when applying $d \gg e^{\epsilon}$.

\section{Data Source of Section \ref{sec:compare}}
\label{sec:data_compare}

We start with the precision. Since HE is equivalent to unbiased CMS+RR with $m = 2$, we use Eq. \eqref{eq:cms_var_rr} to compute its $\widehat{MSE}$, $l_1$ loss, and $l_2$ losses. The computed $\widehat{MSE}$ is the same as in \cite{acharya2019hr}. The $l_1$ and $l_2$ losses of RHR were taken from \cite{chen2020rhr}, but it did not calculate the $\widehat{MSE}$. Section \ref{sec:rhr} shows that RHR is a skewed version of unbiased CMS, and the $\widehat{MSE}$ of RHR could be worse than that from the CMS+RR with $m = 2$ (i.e., HE), so the $\widehat{MSE}$ of RHR is $\Omega\left(\left(\frac{e^{\epsilon} + 1}{e^{\epsilon} - 1}\right)^2\right)$. OLH is equivalent to the unbiased CMS with $m = 1 + e^{\epsilon}$, so we use Eq. \eqref{eq:cms_var_rr} and follow Section \ref{sec:optimized_cms} to calculate its $l_1$ loss, $l_2$ loss, and $\widehat{MSE}$. Section \ref{sec:recursive_cms} shows that CMS+HE is equivalent to the unbiased CMS+RR with $m = 2$ (i.e., HE), so its $l_1$ loss, $l_2$ loss, and $\widehat{MSE}$ are the same as HE. The $l_1$ and $l_2$ losses of SS are presented in \cite{ye2018subset}, but their calculation of $l_1$ is incorrect: (1) they approximate $\sum_x \in [d] |\hat{f}(x) - f(x)|$ as a normal distribution, whereas the central limit theorem requires each $|\hat{f}(x) - f(x)|$ to be i.i.d, and (2) they derived the result based on the Mean Absolute Deviation of the approximated Gaussian distribution, which is equivalent to using $|\sum_x \hat{f}(x) - f(x)|$ to represent $\sum_x |\hat{f}(x) - f(x)|$. The correct way to compute $l_1$ is to use Cauchy–Schwarz inequality, i.e., $l_1 \le \sqrt{d \: l_2}$, which provides a tight upper bound of $l_1$ as  $\frac{2 d e^{\epsilon / 2}}{e^{\epsilon} - 1} $. The $\widehat{MSE}$ of SS is derived using Eq. (2) in \cite{wang2017localhashing} with 

\begin{align*}
        & p = \frac{k e^{\epsilon}}{k e^{\epsilon} + d - k} \\
        & q = \frac{k e^{\epsilon}}{k e^{\epsilon} + d - k} \frac{k-1}{d-1} + \frac{d - k}{k e^{\epsilon} + d - k} \frac{k}{d-1} \\
        & k = d / (e^{\epsilon} + 1) \\
        & f = 1
\end{align*}
\noindent where $k$ represents the size of a subset in SS. $f = 1$ because the $Var(\hat{f}(x))$ increases with $f(x)$ for SS. 

The $\widehat{MSE}$ of a-RP and RP is calculated using Eq. (2) in \cite{wang2017localhashing} as well. Both cases where $f(x) = 0$ and $f(x) = 1$ are calculated, and the worst-case variance is the maximum between them. 

Next, we present the data source regarding communication cost. We first compute the cost of OCMS+RR, which requires the server and a client to sync on the assigned hash function and the perturbed response $z^{(i)}$. Section \ref{sec:imperfect_hashing} shows that $2 \lceil \log_2 \max \{d + 1, 5 m \} \rceil$ bits are required to represent an approximately pairwise independent hash function. Also, we need another $\log_2 m$ bits to sync the perturbed response $z^{(i)}$. Given that OCMS+RR has either $m = 1 + e^{\epsilon / 2}$ or $m = 1 + e^{\epsilon}$ depending on the optimization goal, the overall communication costs of MSE-OCMS+RR and $l$-OCMS+RR are approximately $\max\{ 2\log_2 d + \epsilon/2, \frac{3\epsilon}{2} + 6 \}$ and  $\max\{ 2\log_2 d + \epsilon/2, 3\epsilon + 6 \}$, respectively.

The costs of HE and RHR are taken from their respective literature \cite{acharya2019hr} and \cite{chen2020rhr}. It is notable that we do not use the communication cost from their public-randomness scheme because they fail to consider the cost of syncing the public randomness, as pointed out in Section \ref{sec:public_private_randomness}. For OLH, its hash function maps $[d]$ to $[1 + e^{\epsilon}]$, and each mapping is mutually independent, so it requires $d \log_2 (1 + e^{\epsilon}) \approx d \epsilon$ bits to represent all the mappings. The original CMS \cite{apple_privacy} did not consider pairwise independence, so its hash function mapping from $[d]$ to $[m]$ requires $d \log_2 m$ bits to communicate. SS randomly selects $\frac{d}{e^{\epsilon} + 1}$ values from $[d]$, and this subset selection requires $\frac{d}{e^{\epsilon} + 1}$ bits to represent. RAPPOR generates a one-hot vector with size $d$, which requires $d$ bits to transfer.

\section{Derivation of the Expectation}
\label{sec:expect_proof}

Let's focus on the term $\hat{y}^{(i)}[h_{j^{(i)}}(x)]$ in the definition of $f(x)$ in Eq. \eqref{eq:estimator_def}, which can be expanded as 
\begin{multline}
    \hat{y}^{(i)}[h_{j^{(i)}}(x)] = R(h_{j^{(i)}}(X^{(i)}))[h_{j^{(i)}}(x)]
    \\ = \sum_{j \in [k]} \mathbf{1}\{j = j^{(i)} \} R(h_j(X^{(i)}))[h_{j}(x)] ,
    \label{eq:y_h_j_expand}
\end{multline}

\noindent where $\mathbf{1}\{*\}$ returns one if $*$ is true. Given that $h_{j^{(i)}}$ is uniformly sampled from $\mathcal{H}$, we have
\begin{multline}
    E[ \hat{y}^{(i)}[h_{j^{(i)}}(x)] ] =\sum_{j \in [k]} E \bigl[ \mathbf{1}\{j = j^{(i)} \} \\ R(h_j(X^{(i)}))[h_{j}(x)] \bigr]  = \sum_{j \in [k]} \frac{1}{k} c_j(X^{(i)}, x) , 
    \label{eq:y_h_j_expect}
\end{multline}

\noindent where $c_j(X^{(i)}, x)$ indicates whether $X^{(i)}$ and $x$ collide in the hash function $h_j$, and $E \bigl[ R(h_j(X^{(i)}))[h_{j}(x)] \bigr] = c_j(X^{(i)}, x)$ is derived from Property \ref{pro:ldp_1}. Substituting the above equation into Eq. \eqref{eq:estimator_def}, we have
\begin{multline}
    E[\hat{f}(x)] = \frac{m}{ n (m - 1) } \sum_{i \in [n]} E\bigl[   \hat{y}^{(i)}[h_{j^{(i)}}(x)] \bigr]  - \frac{1}{m - 1}
    \\ = \frac{m}{ n (m - 1) } \sum_{i \in [n]} \sum_{j \in [k]} \frac{ c_j(X^{(i)}, x)}{k} -  \frac{1}{m - 1}
    \\ =  \frac{m}{m - 1} [f(x) + \sum_{x' \in [d] \backslash x} \sum_{j \in [k]}  \frac{c_j(x, x')}{k} f(x')] - \frac{1}{m - 1} , \notag
\end{multline}

\noindent and Theorem $\ref{thm:expect_fx}$ is proved. $\qed$

Substituting $\forall x, x'$, where $x \ne x'$ and $\sum_{j \in [k]} \frac{c_j(x, x')}{k} = \frac{1}{m}$ into the above equation, we have
\begin{multline}
    E[\hat{f}(x)] =  \frac{m}{m - 1} [f(x) + \sum_{x' \in [d] \backslash x} \frac{1}{m} f(x')] - \frac{1}{m - 1} \\
    =  \frac{m}{m - 1} [f(x) + (1 - f(x)) \frac{1}{m}] - \frac{1}{m - 1}  = f(x) . \notag
\end{multline}

Regarding its inverse, whose condition is that there exists an $x'$ such that $\sum_{j \in [k]} \frac{c_j(x, x')}{k} \ne \frac{1}{m}$. If we construct a dataset containing only $x'$ and $x$, then we have
\begin{multline}
    E[\hat{f}(x)] =  \frac{m}{m - 1} [f(x) + (1 - f(x)) \sum_{j \in [k]} \frac{c_j(x, x')}{k}] \\ - \frac{1}{m - 1} 
    \ne    f(x) , \notag
\end{multline}

\noindent so Corollary \ref{cor:fx_no_bias} is proved. $\qed$

\section{Derivation of the Variance}
\label{sec:var_proof}

Let's start with the variance of an individual response $\hat{y}^{(i)}[h_{j^{(i)}}(x)]$:
\begin{multline}
    Var \bigl( \hat{y}^{(i)}[h_{j^{(i)}}(x)] \bigr) 
    \\ = E\bigl[ \bigl( \hat{y}^{(i)}[h_{j^{(i)}}(x)] \bigr)^2  \bigr] - \bar{c}(X^{(i)}, x)^2 ,
    \label{eq:var_step_1}
\end{multline}

\noindent where $E\bigl[ \hat{y}^{(i)}[h_{j^{(i)}}(x)] \bigr] = \bar{c}(X^{(i)}, x)$ given Eq. \eqref{eq:y_h_j_expect}. Then,
\begin{multline}
    E\bigl[ \bigl( \hat{y}^{(i)}[h_{j^{(i)}}(x)] \bigr)^2  \bigr] \\ = \sum_{j \in [k]} E\bigl[ \bigl( \mathbf{1}\{j = j^{(i)} \} R(h_j(X^{(i)}))[h_{j}(x)] \bigr)^2 \bigr]
    \\ =  \sum_{j \in [k]} E[  \mathbf{1}\{j = j^{(i)} \} ] E\bigl[ \bigl( R(h_j(X^{(i)}))[h_{j}(x)] \bigr)^2 \bigr]
    \\ = \frac{1}{k} \sum_{j \in [k]}  E\bigl[ \bigl( R(h_j(X^{(i)}))[h_{j}(x)] \bigr)^2 \bigr]
    \\ \stackrel{\text{(a)}}{=} \frac{1}{k} \sum_{j \in [k]} \biggl( Var\bigl( R(h_j(X^{(i)}))[h_{j}(x)] \bigr) + c_j(X^{(i)}, x) \biggr) \notag
\end{multline}

\noindent where $\stackrel{\text{(a)}}{=}$ holds because $E\bigl[ R(h_j(X^{(i)}))[h_j(x)] \bigr]^2 = c_j(X^{(i)}, x)^2 = c_j(X^{(i)}, x)$. Integrating the above equation into Eq. \eqref{eq:var_step_1}, we have
\begin{multline}
    Var \bigl( \hat{y}^{(i)}[h_{j^{(i)}}(x)] \bigr) 
    = \frac{1}{k} \sum_{j \in [k]}  Var\bigl( R(h_j(X^{(i)}))[h_{j}(x)] \bigr) \\ + \bar{c}(X^{(i)}, x) - \bar{c}(X^{(i)}, x)^2.
    \label{eq:var_step_2}
\end{multline}

Now, we will study the variance of $\sum_{i \in [n]} \hat{y}^{(i)}[h_{j^{(i)}}(x)]$:
\begin{multline}
    Var\bigl( \sum_{i \in [n]} \hat{y}^{(i)}[h_{j^{(i)}}(x)]  \bigr) 
     = \sum_{i \in [n]} Var\bigl(  \hat{y}^{(i)}[h_{j^{(i)}}(x)]  \bigr) \\ + \sum_{i_1 \ne i_2} Cov\bigl(  \hat{y}^{(i_1)}[h_{j^{(i_1)}}(x)], \hat{y}^{(i_2)}[h_{j^{(i_2)}}(x)] \bigr) . \notag
\end{multline}

Focusing on the covariance and expanding $\hat{y}^{(i)}[h_{j^{(i)}}(x)]$ as Eq. \eqref{eq:y_h_j_expand}, we have
\begin{multline}
    Cov\bigl(  \hat{y}^{(i_1)}[h_{j^{(i_1)}}(x)], \hat{y}^{(i_2)}[h_{j^{(i_2)}}(x)] \bigr) 
     = \\ Cov\bigl(  \sum_{j \in [k]} \mathbf{1}\{j = j^{(i_1)} \} R(h_j(X^{(i_1)}))[h_{j}(x)] , \\ \sum_{j \in [k]} \mathbf{1}\{j = j^{(i_2)} \} R(h_j(X^{(i_2)}))[h_{j}(x)] \bigr) . \notag
\end{multline}

Given that $Cov(X, Y) = E[XY] - E[X]E[Y]$, we calculate:
\begin{multline}
    E\bigl[  \sum_{j_1 \in [k]} \sum_{j_2 \in [k]} \mathbf{1}\{j_1 = j^{(i_1)} \} \mathbf{1}\{j_2 = j^{(i_2)} \} \\  R(h_{j_1}(X^{(i_1)}))[h_{j_1}(x)]   R(h_{j_2}(X^{(i_2)}))[h_{j_2}(x)] \bigr]
    \\ \stackrel{\text{(a)}}{=} \sum_{j_1 \in [k]} \sum_{j_2 \in [k]} E[   \mathbf{1}\{j_1 = j^{(i_1)} \}] E[\mathbf{1}\{j_2 = j^{(i_2)} \}] \times \\  E[R(h_{j_1}(X^{(i_1)}))[h_{j_1}(x)]]   E[R(h_{j_2}(X^{(i_2)}))[h_{j_2}(x)]] 
    \\ = \sum_{j_1 \in [k]} \sum_{j_2 \in [k]} \frac{1}{k^2} c_{j_1}(X^{(i_1)}, x) c_{j_2}(X^{(i_2)}, x) \\ = \bar{c}(X^{(i_1)}, x) \bar{c}(X^{(i_2)}, x) ,
    \label{eq:var_cov_step_1}
\end{multline}

\noindent where $\stackrel{\text{(a)}}{=}$ holds because the assignment of a hash function is simply a uniform sampling, which is independent of everything, and $R(h_{j_1}(X^{(i_1)}))[h_{j_1}(x)]$ is also independent of $R(h_{j_2}(X^{(i_2)}))[h_{j_2}(x)]$ given Property \ref{pro:ldp_2}.

Utilizing Eq. \eqref{eq:y_h_j_expect} and \eqref{eq:var_cov_step_1}, we have 
\begin{multline}
    Cov\bigl(  \hat{y}^{(i_1)}[h_{j^{(i_1)}}(x)], \hat{y}^{(i_2)}[h_{j^{(i_2)}}(x)] \bigr)
    \\ = \bar{c}(X^{(i_1)}, x) \bar{c}(X^{(i_2)}, x) - \bar{c}(X^{(i_1)}, x) \bar{c}(X^{(i_2)}, x) = 0 . \notag
\end{multline}

\noindent As a result, we have
\begin{multline}
    Var\bigl( \sum_{i \in [n]} \hat{y}^{(i)}[h_{j^{(i)}}(x)]  \bigr) 
     = \sum_{i \in [n]} Var\bigl(  \hat{y}^{(i)}[h_{j^{(i)}}(x)]  \bigr)
     \\ \stackrel{\text{Eq. \eqref{eq:var_step_2}}}{=} \sum_{i \in [n]} \biggl[ \biggl( \frac{1}{k} \sum_{j \in [k]}  Var\bigl( R(h_j(X^{(i)}))[h_{j}(x)] \bigr) \biggr) \\ +  \bar{c}(X^{(i)}, x) - \bar{c}(X^{(i)}, x)^2 \biggl]
     \\ = n \sum_{x' \in [d]} \biggl[ \biggl( \frac{1}{k} \sum_{j \in [k]}  Var\bigl( R(h_j(x'))[h_{j}(x)] \bigr) \biggr) \\ +  \bar{c}(x', x) - \bar{c}(x', x)^2  \biggr]  f(x') . \notag
\end{multline}

\noindent and subsequently, we have
\begin{multline}
    Var( \hat{f}(x) ) = \frac{m^2}{(m - 1)^2 n} \sum_{x' \in [d]} \biggl[ \biggl( \frac{1}{k} \sum_{j \in [k]} \\  Var\bigl( R(h_j(x'))[h_{j}(x)] \bigr) \biggr)  +  \bar{c}(x', x) - \bar{c}(x', x)^2  \biggr]  f(x') ,
\end{multline}

\noindent which proves Eq. \eqref{eq:var_hat_f}. $\qed$ 

When considering Property \ref{pro:ldp_3}, we have $Var\bigl( R(h_j(x'))[h_{j}(x)] \bigr) = Var(R|=)$ if $h_j(x') = h_j(x)$. Otherwise, it equals $Var(R|\ne)$. When $x' = x$, they will collide in every hash function, so $\bar{c}(x', x) = 1$ and $Var\bigl( R(h_j(x'))[h_{j}(x)] \bigr) = Var(R|=)$ in this case.  On the other hand, when $x' \ne x$, we have
\begin{multline}
     \frac{1}{k} \sum_{j \in [k]}  Var\bigl( R(h_j(x'))[h_{j}(x)] \bigr) \\ = \frac{1}{k} \sum_{j \in [k]} [  c_j(x', x) Var(R|=)  +  (1 - c_j(x', x)) Var(R | \ne)] 
     \\ =  \bar{c}(x', x) Var(R|=) +   (1 - \bar{c}(x', x)) Var(R | \ne).
     \label{eq:var_sym_step_1}
\end{multline}

Therefore, when substituting Eq. \eqref{eq:var_sym_step_1} for $x' \ne x$ and the aforementioned parameters with $x = x'$ into Eq. \eqref{eq:var_hat_f}, we derived Eq. \eqref{eq:var_hat_f_sym}. If $\bar{c}(x', x)  = \frac{1}{m}$, it becomes Eq. \eqref{eq:var_hat_f_pairwise}. $\qed$

\section{Proof of Theorem \ref{thm:cms_rr_bound}}
\label{sec:proof_cms_rr_bound}

Recall the definition of $\hat{f}(x)$ in Eq. \eqref{eq:estimator_def}, which is equivalent to the sum of bounded random variables. Define $Y^{(i)} = \frac{m}{m - 1} \hat{y}^{(i)}[h_j^{(i)}(x)] - \frac{1}{m - 1}$, and we have $n \hat{f}(x) = \sum_i Y^{(i)}$. Note that $\frac{Y^{(i)} - A}{B - A}$ is a Bernoulli random variable where
\begin{align*}
    &  A = \min \hat{f}(x) = \frac{m}{m - 1} \frac{e^{-1}}{e^{\epsilon} - 1} - \frac{1}{m - 1} \\ 
    & B  = \max \hat{f}(x)  = \frac{m}{m - 1} \frac{e^{\epsilon} + m - 2}{e^{\epsilon} - 1} - \frac{1}{m - 1} .
\end{align*}

\noindent Here, we assume that CMS uses RR to perturb the hashed values. Also note that we only consider unbiased CMS, i.e., $E[\hat{f}(x)] = f(x)$. Using the Chernoff (upper) bound, we have $\forall \delta \in [0, 1]$
\begin{multline}
    Pr[ \sum_i \frac{Y^{(i)} - A}{B - A} \ge (1 + \delta) n \frac{f(x) - A}{B - A} ] \\ \le \exp(- \frac{n}{3} \frac{f(x) - A}{B - A} \delta^2) .
    \label{eq:proof_cms_rr_bound_step_1}
\end{multline}

Define $\delta = \frac{\alpha \sqrt{Var(\hat{f}(x))}}{f(x) - A}$. Given that  $\forall \delta \in [0, 1]$, it is equivalent to requiring $\alpha \in [0, \sqrt{\frac{n e^{\epsilon}}{m - 1}}]$. Thus, Eq. \eqref{eq:proof_cms_rr_bound_step_1} can be rewritten as
\begin{multline}
     Pr[ \hat{f}(x) \ge f(x) + \alpha \sqrt{Var(\hat{f}(x))} ] \\ \le \exp(- \frac{n}{3} \frac{Var(\hat{f}(x))}{(B - A) (f(x) - A)} \alpha^2) \notag
\end{multline}

$Var(\hat{f}(x))$  is derived in Eq. \eqref{eq:cms_var_rr}, and $ \frac{Var(\hat{f}(x))}{(B - A) (f(x) - A)} $ decreases with $f(x)$ (one can verify it using derivatives). Calculating the minimum at $f(x) = 1$, we have

\begin{equation*}
    \frac{Var(\hat{f}(x))}{(B - A) (f(x) - A)} \ge \frac{m - 1}{e^{\epsilon} + m - 1} .
\end{equation*}

\noindent As a result,
\begin{multline}
     Pr[ \hat{f}(x) \ge f(x) + \alpha \sqrt{Var(\hat{f}(x))} ] \\ \le \exp(- \frac{n}{3} \frac{m - 1}{e^{\epsilon} + m - 1} \alpha^2) .\notag
\end{multline}

The same proof is also applicable to $\hat{f}(x) \le f(x) - \alpha \sqrt{Var(\hat{f}(x))}$. Therefore,
\begin{multline}
     Pr[ | \hat{f}(x) - f(x)   | \ge  + \alpha \sqrt{Var(\hat{f}(x))} ] \\ \le 2 \exp(- \frac{n}{3} \frac{m - 1}{e^{\epsilon} + m - 1} \alpha^2) . \qed \notag
\end{multline}

\section{Proof regarding Preferring RR to RAPPOR in CMS}
\label{sec:proof_optimized_cms_rappor}

The variance of symmetry and asymmetry RAPPOR is listed below:
\begin{equation}
  \forall a,b :  Var( sRP(a)[b] ) = \frac{e^{\epsilon / 2}}{ ( e^{\epsilon / 2} - 1 )^2 }  . \label{eq:var_srp}
\end{equation}
\begin{multline}
    Var( aRP(a)[b] ) = \\ \frac{1}{ ( e^{\epsilon} - 1 )^2 } 
    \begin{cases}
        (e^{\epsilon} + 1)^2 & \text{ if } a = b\\
        4 e^{\epsilon}  & \text{ if } a \ne b ,
    \end{cases}
    \label{eq:var_arp}
\end{multline}

Since their $Var(a)[a]$ and $Var(a)[b]$ are independent of $m$, the corresponding $Var(\hat{f}(x))$ decreases when $m$ increases. Thus, when $m$ is large enough, we have
\begin{multline}
    Var(\hat{f}(x)_{RP}) \rightarrow \frac{1}{n} [  Var(R | \ne) (1 - f(x))  +  \\  Var(R | =) f(x) ].
    \label{eq:rp_large_m}
\end{multline}

Following Appendix \ref{sec:proof_optimized_cms_rr} and given unbiased CMS, the worst-case MSE of symmetry and asymmetry RAPPOR are identical to their variance, which are $ \frac{e^{\epsilon/2}}{n(e^{\epsilon/2} - 1)^2}$ and $ \frac{f^*(e^{\epsilon}-1)^2 + 4^{\epsilon}}{n(e^{\epsilon} - 1)^2}$, respectively.

For asymmetric RAPPOR, we realize $Var(\hat{f}(x)_{aRP}) > Var^*(\hat{f}(x)_{RR})$ when $f^* > 0$, so the CMS with asymmetric RAPPOR cannot serve as the optimized CMS. For symmetric RAPPOR, $Var(\hat{f}(x)_{sRP}) > Var^*(\hat{f}(x)_{RR})$ when $0 \le f^* < \frac{1}{2}$, and $Var(\hat{f}(x)_{sRP}) = Var^*(\hat{f}(x)_{RR})$ when $\frac{1}{2} \le f^* \le 1$. Thus, the CMS with symmetric RAPPOR is also ruled out. As a result, The choice of RR is preferred.

Similar to Appendix \ref{sec:proof_optimized_cms_rr}, we can use the variance to derive $l_2$ loss, which are $ \frac{d e^{\epsilon/2}}{n(e^{\epsilon/2} - 1)^2}$ and $\frac{4 d e^{\epsilon}}{n (e^{\epsilon} - 1)^2}$ for symmetric and asymmetric RAPPOR, respectively. we observe that CMS with symmetric RAPPOR will always have a larger $l_2$ loss than the CMS+RR. For asymmetric RAPPOR, its $l_2$ loss is the same as that of CMS+RR only when $d$ is large enough. However, its communication cost is too high because RAPPOR requires sending a vector of $m$ bits to the server, and Eq. \eqref{eq:rp_large_m} requires $m \rightarrow \infty$. To determine the necessary value of $m$, we solve the following equation for asymmetric RAPPOR:

\begin{equation}
     l_2(\hat{f}_{aRP}) = d Var(\hat{f}(x)_{aRP}) = (1 + \tau)  \frac{4 d e^{\epsilon}}{(e^{\epsilon} - 1)^2} , \notag
\end{equation}

\noindent where $\tau$ is a small number like 0.01. $m$ is solved as
\begin{equation}
    m = \Omega(\frac{(e^{\epsilon} - 1)^2}{(1 + \tau) e^{\epsilon}}) .
    \label{eq:rappor_bits}
\end{equation}

On the other hand, CMS+RR only requires $\log_2(1 + e^{\epsilon})$ bits to output the perturbed result. Thus, CMS+RR is always preferred when optimizing $l_2$ loss due to its low communication cost. 

\section{Proof regarding Imperfect Hashing}
\label{sec:proof_imperfect_hashing}

Let's start by proving Theorem \ref{thm:var_g_x_approach_f_x}. Observe $\frac{\partial Var( g(x \mid \mathcal{H}_{api}) )}{\partial m'}$, and note that it is linear to $f(x)$. Considering the derivative at $f(x) = 1$ and $f(x) = 0$ yields:
\begin{multline}
    - \max\{\frac{2m' Var(R|=)}{n(m'-1)^3} , \\ \frac{(m' - 1) (Var(R | \ne) + 1) + (m' + 1) Var(R|=)}{n(m'-1)^3}  \} \\ \le \frac{\partial Var( g(x  | \mathcal{H}_{api}) )}{\partial m'} < 0 .
\end{multline}

\noindent  Observing that $Var(R|=) \ge Var(R|\ne)$ is satisfied by the LDP mechanism such as randomized response and RAPPOR, we have

\begin{equation}
    - \frac{2m' Var(R|=) + m' - 1}{n(m'-1)^3} \le \frac{\partial Var( g(x  | \mathcal{H}_{api}) )}{\partial m'} < 0 .
\end{equation}

At the same time, we have $m' = m - \frac{m}{(2q + 1)^2}$. Denote $Var( \hat{f}(x) )$ as the variance of the CMS using perfect hashing. We want $Var( g(x  | \mathcal{H}_{api}) )$ to approach $Var( \hat{f}(x) )$, which is formulated as $Var( g(x  | \mathcal{H}_{api}) ) \le Var( \hat{f}(x) ) (1 + \tau)$, where $\tau$ is a small number like 0.01. This can be translated to 

\begin{equation}
    \frac{2m Var(R|=) + m - 1}{n(m-1)^3} \frac{m}{(2q + 1)^2} \le \tau Var( \hat{f}(x) ) , \notag
\end{equation}

\noindent which can be organized as
\begin{equation}
    \frac{2m Var(R|=) + m - 1}{n(m-1)^3} \frac{m}{ \tau Var( \hat{f}(x) ) } \le  (2q + 1)^2 , 
    \label{eq:var_g_x_approach_f_x_condition}
\end{equation}

\noindent thus proving Theorem \ref{thm:var_g_x_approach_f_x}. When $m = 1 + e^{\epsilon /2}$, $Var( \hat{f}(x) )$ becomes $\frac{e^{\epsilon / 2}}{(e^{\epsilon / 2} - 1)^2}$, and $\frac{Var(R|=)}{ Var( \hat{f}(x))}$ decreases with $\epsilon$. Thus, we can substitute $\epsilon = 0$ into Eq. \eqref{eq:var_g_x_approach_f_x_condition}, and we have $2 q + 1 > \sqrt{\frac{1}{\tau}}$. The MSE part of Corollary \ref{cor:g_close_to_f} is proved. $\qed$

If $\forall x: Var(g(x)) \le (1 + \tau) Var(\hat{f}(x))$, then $l_2(g) \le (1 + \tau) l_2(\hat{f})$. When $m = 1 + e^{\epsilon}$, $Var(\hat{f}(x))$ increases with $f(x)$, so we have  $ Var( \hat{f}(x) | f(x) = 0) \le Var( \hat{f}(x) )$. Thus, if

\begin{multline}
    \frac{2m Var(R|=) + m - 1}{n(m-1)^3} \frac{m}{ \tau Var( \hat{f}(x) | f(x) = 0) } \\ \le  (2q + 1)^2 
    \label{eq:l2_g_x_approach_f_x_condition_step_1}
\end{multline}

\noindent is satisfied, Eq. \eqref{eq:var_g_x_approach_f_x_condition} will also be satisfied. Substituting $\frac{4 e^{\epsilon}}{(e^{\epsilon} - 1)^2}$ into $Var(\hat{f}(x) | f(x) = 0)$ in Eq. \eqref{eq:l2_g_x_approach_f_x_condition_step_1} and setting $m =  1 + e^{\epsilon}$, we also have the left-hand side of Eq. \eqref{eq:l2_g_x_approach_f_x_condition_step_1} decreasing with $\epsilon$. Since this inequality is still satisfied when $\epsilon = 0$, it derives $2 q + 1 > \sqrt{\frac{1}{\tau}}$. Thus, The $l_2$ part of Corollary \ref{cor:g_close_to_f} is proved. $\qed$

\section{Proof of Theorem \ref{thm:exp_and_var_of_bias}}
\label{sec:proof_exp_and_var_of_bias}

Consider each $c_j(x, x')$ as an independent Bernoulli random variable with probability $\frac{1}{m}$ of being one. Given that $E[c_j(x, x')] = \frac{1}{m}$ and $Var(c_j(x, x')) = \frac{m - 1}{m}$, we have
\begin{multline}
    \mathop{E}_{ \forall c_j(x, x') } [ E[\hat{f}(x)] ] \\ =  \frac{m}{m - 1} [f(x) + \sum_{x' \in [d] \backslash x} \frac{k}{k m} f(x')] -  \frac{1}{m - 1} \\
    =  \frac{m}{m - 1} [f(x) + (1 - f(x)) \frac{1}{m}] - \frac{1}{m - 1}  = f(x) , \notag
\end{multline}

\noindent and
\begin{multline}
    \mathop{Var}_{ \forall c_j(x, x') } [ E[\hat{f}(x)] ] =  \frac{m^2}{(m - 1)^2} \biggl( \sum_{x' \in [d] \backslash x} \frac{m - 1}{m^2 k} f(x')^2 \biggr) \\ = \frac{1}{(m - 1)k} \sum_{x' \in [d] \backslash x} f(x')^2  . \notag 
\end{multline}

\noindent Thus, Theorem \ref{thm:exp_and_var_of_bias} is proved. $\qed$

\section{Decoding Hadamard Encoding}
\label{sec:decode_hardmard}

Define $H$ to be a scaled Walsh Hadamard matrix, where $H[i, j] = (-1)^{i \cdot j}$ with $\cdot$ representing bitwise multiplication. Hadamard encoding employs a LDP mechanism to perturb $H [X^{(i)} + 1, j^{(i)}]$ of each object, where $X^{(i)}$ and $j^{(i)}$ are as defined in  Section \ref{sec:revisit_cms}. \cite{bassily2015hadamard} has proved that 

\begin{equation}
    \hat{f}_H(x) = \sum_{i \in [n]} \hat{H}[X^{(i)} + 1, j^{(i)}] H[j^{(i)}, x + 1] \notag
\end{equation}

\noindent unbiasedly estimates the frequency of $f(x)$. Here, $\hat{H}[X^{(i)} + 1, j^{(i)}] = R( H [X^{(i)} + 1, j^{(i)}] )$, with $R$ denoting the LDP reconstruction process as detailed in Section \ref{sec:frequency_estimation}. 

Interpreting $H[j^{(i)}, :]$ as a hash function, $\hat{H}[X^{(i)} + 1, j^{(i)}] H[j^{(i)}, x + 1]$ yields +1 if both $X^{(i)}$ and $x$ hash to the same value, and -1 otherwise. This behavior is analogous to $\hat{y}^{(i)}[h_{j^{i}}(x)]$ in Eq. \eqref{eq:estimator_def}. However, $\hat{y}^{(i)}[h_{j^{i}}(x)]$ returns zero when $X^{(i)}$ and $x$ hash to different values. By considering the constants $\frac{m}{m - 1}$ and $\frac{1}{m - 1}$ in Eq. \eqref{eq:estimator_def}, and setting $m = 2$, we derive $2 \hat{y}^{(i)}[h_{j^{i}}(x)] - 1$, which produces results identical to $ \hat{H}[X^{(i)} + 1, j^{(i)}] H[j^{(i)}, x + 1]$. Consequently, the decoding process for Hadamard encoding is equivalent to Eq. \eqref{eq:estimator_def}.

\section{Upper Bound of the worst-case MSE estimator}
\label{sec:upper_bound_mse}

The upper bound of $\overline{MSE}(\hat{f})$ is set to be $[1 + \frac{2}{t} \bigl(\sqrt{t \log(20 |\mathbf{x}|)} + \log(20 |\mathbf{x}|)\bigr)] V $ based on the following theorem:

\begin{theorem}
If $\forall x \in \mathbf{x}: E[(\hat{f}(x) - f(x))^2] \le V$, where $V$ is constant, we have
\begin{multline*}
    Pr \bigl( \overline{MSE}(\hat{f}) \ge [1 + \frac{2}{t} \bigl(\sqrt{t \log(20 |\mathbf{x}|)} + \\ \log(20 |\mathbf{x}|)\bigr)] V \bigr) \le 0.05 ,
\end{multline*}
where $t$ is the experiment rounds.
\end{theorem}

\noindent Proof: Similar to Appendix \ref{sec:proof_cms_rr_bound}, we define $Y^{(i)} = \frac{m}{m - 1} \hat{y}^{(i)}[h_j^{(i)}(x)] - \frac{1}{m - 1}$, and $\frac{Y^{(i)} - A}{B - A}$ is a Bernoulli random variable (see Appendix \ref{sec:proof_cms_rr_bound} for the definition of $A$ and $B$). All the $X^{(i)}$ can be placed into two groups based on whether $X^{(i)} = x$. If $X^{(i)} = x$, $\frac{Y^{(i)} - A}{B - A} \sim \text{Ber}\left(\frac{e^{\epsilon}}{e^{\epsilon} + m - 1}\right)$, where $\text{Ber}(p)$ denotes the Bernoulli distribution with probability $p$. On the other hand, if $X^{(i)} \ne x$, $\frac{Y^{(i)} - A}{B - A} \sim \text{Ber}\left(\frac{1}{m} \frac{e^{\epsilon}}{e^{\epsilon} + m - 1} + \frac{m - 1}{m} \frac{1}{e^{\epsilon} + m - 1}\right)$. For convenience, we denote $p_1 = \frac{e^{\epsilon}}{e^{\epsilon} + m - 1}$ and $p_2 = \frac{1}{m} \frac{e^{\epsilon}}{e^{\epsilon} + m - 1} + \frac{m - 1}{m} \frac{1}{e^{\epsilon} + m - 1}$. 

Subsequently, $\frac{n (\hat{f}(x) - A)}{B - A}$ is equivalent to the sum of two binomial random variables, where the first is sampled from $\text{BN}(n f(x), p_1)$ and the second is sampled from $\text{BN}(n (1 - f(x)), p_2)$, where $\text{BN}(n, p)$ denotes the binomial distribution with $n$ trials each having a success probability $p$.  

When $n$ is large enough, both binomial random variables can be approximated as Gaussian random variables, which are sampled from $\mathcal{N}(n f(x) p_1, n p_1 (1 - p_1))$ and $\mathcal{N}(n (1 - f(x)) p_2, n p_2 (1 - p_2))$ respectively, where $\mathcal{N}(\mu, \sigma^2)$ denotes a Gaussian distribution with mean $\mu$ and variance $\sigma^2$. Since the sum of these two Gaussian random variables is also a Gaussian random variable, $\hat{f}(x)$ is approximated as
\begin{multline*}
    \frac{n (\hat{f}(x) - A)}{B - A} \sim \mathcal{N}(n f(x) p_1 + n (1 - f(x)) p_2, \\ n f(x) p_1 (1 - p_1) +   n (1 - f(x)) p_2 (1 - p_2) ).
\end{multline*}

\noindent One can verify that the above equation is equivalent to
\begin{equation*}
    \frac{n (\hat{f}(x) - A)}{B - A} \sim \mathcal{N}(\frac{n (f(x) - A)}{B - A}, \frac{n^2 Var(\hat{f}(x))}{(B - A)^2} ),
\end{equation*}

\noindent which can be refactored as

\begin{equation*}
    \frac{\hat{f}(x)  - f(x)}{\sqrt{Var(\hat{f}(x))}} \sim \mathcal{N}(0,  1 ).
\end{equation*}

Thus, $\frac{\hat{f}(x) - f(x)}{\sqrt{Var(\hat{f}(x))}}$ of each experiment can be considered a standard Gaussian random variable. There are $t$ rounds of the experiment; using the Laurent-Massart bound \cite{laurent2000adaptive}, we have
 \begin{multline*}
     Pr \biggl(\sum_t [(\frac{\hat{f}(x)_t  - f(x)}{\sqrt{Var(\hat{f}(x))}})^2 - 1]  \\ \ge 2 (\sqrt{t \alpha} + \alpha) \biggr) \le e^{-\alpha} .
 \end{multline*}

\noindent Define $S(x) = \frac{1}{t} (\hat{f}(x)_t - f(x))^2$, and we have

\begin{equation*}
     Pr \biggl(S(x)  \ge [1 + \frac{2}{t} (\sqrt{t \alpha} + \alpha)] Var(\hat{f}(x)) \biggr) \le e^{-\alpha} . 
\end{equation*}

Given our assumption that $E[(\hat{f}(x) - f(x))^2] = Var(\hat{f}(x)) \le V$ is identical for $\forall x \in \mathbf{x}$, we have

\begin{equation*}
     Pr \biggl(S(x)  \ge [1 + \frac{2}{t} (\sqrt{t \alpha} + \alpha)] V \biggr) \le e^{-\alpha} . 
\end{equation*}

Note that $\forall x: S(x) < (1 + \frac{2}{t} (\sqrt{t \alpha} + \alpha)) V$ is equivalent to $\max_x S(x) < (1 + \frac{2}{t} (\sqrt{t \alpha} + \alpha)) V$. Given that $\overline{MSE}(\hat{f}) = \max_x S(x)$, we have

\begin{equation}
    Pr \bigl( \overline{MSE}(\hat{f}) < [1 + \frac{2}{t} (\sqrt{t \alpha} + \alpha)] V \bigr)  > (1 - e^{-\alpha})^{|\mathbf{x}|} ,
    \label{eq:uppeer_bound_mse_step_1}
\end{equation}

\noindent which decreases with $|\mathbf{x}|$. Even when $|\mathbf{x}| \rightarrow \infty$, if $ e^{-\alpha} |\mathbf{x}| = \frac{1}{20}$, we still have $(1 -  e^{-\alpha})^{|\mathbf{x}|} \approx 0.95$. Thus, we have $\alpha = \log(20 |\mathbf{x}|)$. Then, we obtain

\begin{multline*}
    Pr \bigl( \overline{MSE}(\hat{f}) < [1 + \frac{2}{t} \bigl(\sqrt{t \log(20 |\mathbf{x}|)} + \\ \log(20 |\mathbf{x}|)\bigr)] V \bigr)  > 0.95 . \qed
\end{multline*}

\end{document}